\documentclass[a4paper]{article}
\usepackage{a4wide}
\usepackage[UKenglish]{babel}
\usepackage{graphicx,subfigure,caption}
\graphicspath{{./}{figs/}}
\usepackage[colorlinks=true,linkcolor=blue,citecolor=red]{hyperref}
\usepackage{xcolor}
\usepackage{amssymb,amsthm,empheq,bbold}
\usepackage[inline]{enumitem}
\usepackage{authblk}
\usepackage{calrsfs}
    \DeclareMathAlphabet{\pazocal}{OMS}{zplm}{m}{n}
\usepackage{accents}

\DeclarePairedDelimiter\abs{\lvert}{\rvert}
\DeclarePairedDelimiter\ave{\langle}{\rangle}
\newcommand{\cC}{\mathcal{C}}
\newcommand{\cd}{\mathfrak{d}}
\newcommand{\dWinf}{\underline{\cd}_W}
\newcommand{\cE}{\mathcal{E}}
\DeclareMathOperator*{\esup}{ess\,sup}

\newcommand{\Lap}{\mathcal{L}}
\newcommand{\Leb}{\pazocal{L}^1}
\newcommand{\Lip}{\operatorname{Lip}}
\newcommand{\cm}{\mathfrak{m}}
\newcommand{\N}{\mathbb{N}}
\newcommand{\norm}[2]{\Vert#1\Vert_{#2}}
\newcommand{\cP}{\mathcal{P}}
\newcommand{\pp}{\mathfrak{p}}
\newcommand{\pr}[1]{{}^\prime\!#1}
\newcommand{\qq}{\mathfrak{q}}
\newcommand{\R}{\mathbb{R}}
\newcommand{\cR}{\pazocal{R}}
\newcommand{\cS}{\mathfrak{S}}
\newcommand{\cStilde}{\underset{\smash{\raisebox{0.4ex}{$\scriptstyle\sim$}}}{\cS}}
\newcommand{\ubar}[1]{\underaccent{\bar}{#1}}
\newcommand{\cW}{\pazocal{W}}
\newcommand{\Winf}{\ubar{W}}

\newtheorem{assumption}{Assumption}[section]
\newtheorem{proposition}[assumption]{Proposition}
\newtheorem{theorem}[assumption]{Theorem}
\newtheorem{lemma}[assumption]{Lemma}
\newtheorem{corollary}[assumption]{Corollary}
\theoremstyle{remark}\newtheorem{remark}[assumption]{Remark}
\allowdisplaybreaks

\title{Connectivity and the equilibration of Boltzmann-type equations on dense graphs}
\author{Andrea Tosin}
\affil{{\small Department of Mathematical Sciences ``G. L. Lagrange'', Politecnico di Torino, Italy}}
\date{}
		
\begin{document}
\maketitle
	
\begin{abstract}
We investigate the equilibration of solutions to a homogeneous Boltzmann-type equation whose collision kernel is given by a graphon encoding the connectivity structure of interacting agents. For a broad class of Lipschitz-continuous binary interaction rules, we show that, under suitable conditions relating the interaction parameters to the connectivity properties of the graphon, solutions converge exponentially fast towards those of the classical homogeneous Boltzmann-type equation without preferential connections, and hence towards its equilibrium distributions. We then consider in detail compromise-diffusion interaction rules and establish a quasi-invariant limit leading to a Fokker--Planck-type equation parametrised by the graphon connectivity. This limiting equation allows us to characterise explicitly both connectivity-independent and connectivity-dependent equilibrium distributions.

\medskip


\medskip

\end{abstract}

\section{Introduction}
\label{sect:intro}
Modern applications of Boltzmann kinetic theory to sociophysics and econophysics, with opinion formation and wealth distribution as prototypical examples, concern multi-agent systems in which interactions are typically \textit{some-to-some} rather than \textit{all-to-all}, as it happens for molecular collisions in classical gas dynamics. This is because agents in such systems exhibit preferential \textit{patterns of connectivity}, which restrict their interactions to other agents with whom they are directly connected. Despite this feature, sociophysical and econophysical systems have been extensively and successfully modelled by means of homogeneous Boltzmann-type equations~\cite{boudin2012PHYSA,boudin2016PHYSA,cordier2005JSP,pareschi2013BOOK,toscani2006CMS}, without accounting for the \textit{network} of connections among agents.

It is only relatively recently that models explicitly incorporating such connection networks into kinetic equations for sociophysical multi-agent systems have been proposed, primarily in the context of opinion formation problems on social networks~\cite{burger2025SIADS,duering2024JNS,fraia2026KRM,loy2022PTRSA,nurisso2026EJAM,toscani2018PRE} or in studies of the spread of infectious diseases~\cite{ali2026PREPRINT,bondesan2026PREPRINT} and population migration across cities~\cite{taralli2026PREPRINT}. See also~\cite{burger2021VJM,burger2022KRM}. Other contributions that can be placed within this framework concern models of brain networks that use mesoscopic kinetic-type equations to describe the progression of Alzheimer's disease~\cite{bertsch2017MMB,bertsch2017JPA} or the activation and deactivation of neurons underlying electrical brain activity~\cite{bisi2026,conte2024SEMA-SIMAI}.

In~\cite{taricco2026PREPRINT} we derived a general homogeneous Boltzmann-type equation describing \textit{networked binary interactions} within a population of agents. Each agent is characterised by a scalar microscopic state $v\in\R$, generically referred to as a \textit{trait}, which is exchanged during interactions, and by a \textit{latent variable} $x\in [0,\,1]$, which accounts for the agent's ``position'' in the connection network. Such a network is assumed to be fixed, i.e. non-coevolving, and is modelled as a \textit{dense graph}, namely the limit, in an appropriate graph-theoretic sense~\cite{lovasz2006JCTB}, of a sequence of finite graphs with an increasing number of vertices. The dense graph is described analytically by a \textit{graphon}, i.e. a measurable symmetric function $W:[0,\,1]^2\to [0,\,1]$ such that $W(x,x_\ast)$ gives the probability that two agents occupying positions $x,\,x_\ast\in [0,\,1]$ in the connection network are connected and can therefore interact. The graphon arises as the limit, with respect to a suitable metric known as the cut distance~\cite{janson2013NYJM}, of a sequence of step functions associated with the (possibly weighted) adjacency matrices of the finite graphs converging to the dense graph.

In~\cite{taricco2026PREPRINT}, the graph-based Boltzmann-type equation was derived as the dense-graph limit of a sequence of corresponding equations defined on the aforementioned finite graphs. In this equation, the graphon $W$ plays the role of the collision kernel, thereby determining the frequency of interactions between agents according to the connectivity structure of the interaction network. The equation takes the form
\begin{equation}
    \frac{\partial f}{\partial t}=\int_0^1\int_\R W(x,x_\ast)\ave*{\frac{1}{J}f(x,\pr{v},t)f(x_\ast,\pr{v}_\ast,t)-
        f(x,v,t)f(x_\ast,v_\ast,t)}\,dv_\ast\,dx_\ast,
    \label{eq:Boltz.graph-strong}
\end{equation}
where the distribution function $f$ is such that $f(x,v,t)\,dx\,dv$ gives the probability that, at time $t>0$, a representative agent of the system occupies a position in $[x,\,x+dx]$ within the interaction network and has a trait in $[v,\,v+dv]$. A graphon concentrated primarily on pairs $(x,\,x_\ast)\in [0,\,1]^2$ for which $x$ and $x_\ast$ take similar values describes a homophilic society. Conversely, one concentrated primarily on pairs $(x,\,x_\ast)\in [0,\,1]^2$ for which $x$ and $x_\ast$ differ substantially may describe a hierarchical society. Between these two somewhat extreme cases lies a wide range of connection patterns, each of which can be represented by a suitable graphon. The latent variable $x$ can, in turn, admit different interpretations depending on the specific application. For instance, in social networks, it may represent a measure of a user's number of connections; in collaboration networks, an agent's expertise within a certain disciplinary spectrum; and in preference networks, a consumer's personal taste.

In this paper, we investigate how the connectivity of the connection network affects the trend to equilibrium of solutions to~\eqref{eq:Boltz.graph-strong} for broad classes of graphons and binary interaction rules. Throughout the paper, we consider a generic symmetric graphon $W$ and define its associated \textit{degree function} by
\begin{equation}
    \cd_W(x):=\int_0^1W(x,x_\ast)\,dx_\ast,
    \label{eq:degree}
\end{equation}
along with the quantities
\begin{equation}
    \dWinf:=\inf_{x\in [0,\,1]}\cd_W(x), \qquad \norm{\cd_W}{\infty}:=\sup_{x\in [0,\,1]}\cd_W(x),
    \label{eq:deg.param}
\end{equation}
which we refer to as \textit{connectivity parameters}. We show that the interplay between $\dWinf$, $\norm{\cd_W}{\infty}$, and the parameters characterising the binary interaction rules governs the equilibration of solutions to~\eqref{eq:Boltz.graph-strong}. In particular, it determines whether these solutions remain close in time to those of the \textit{corresponding graph-independent Boltzmann-type equation} and eventually converge to its equilibrium distribution. By the latter equation, we mean the one using the same binary interaction rules \textit{in the absence of a connection network}:
\begin{equation}
    \frac{\partial F}{\partial t}=\int_\R\ave*{\frac{1}{J}F(\pr{v},t)F(\pr{v}_\ast,t)-F(v,t)F(v_\ast,t)}\,dv_\ast,
    \label{eq:Boltz-strong}
\end{equation}
where $F$ now denotes a distribution function such that $F(v,t)\,dv$ is the probability that, at time $t>0$, a representative agent of the system has a trait in $[v,\,v+dv]$. Indeed,~\eqref{eq:Boltz-strong} is the homogeneous Boltzmann-type equation traditionally used to model sociophysical and econophysical systems under the assumption of all-to-all interactions~\cite{pareschi2013BOOK}.

Regarding the interaction rules, in Section~\ref{sect:struct_asympt.graph-based} we consider general Lipschitz-continuous functions expressing the post-interaction trait of a representative agent in terms of the pre-interaction traits of two randomly selected interacting agents:
$$ v'=\Psi(v,v_\ast). $$
This extends the framework developed in~\cite{taricco2026PREPRINT}, which was restricted to linear interaction rules, to the non-linear setting. More precisely, letting $p,\,q\geq 0$ denote the (possibly stochastic) Lipschitz constants of the interaction rule, we prove that, under the condition
\begin{equation}
    \ave{p+q}<\frac{\dWinf}{\norm{\cd_W}{\infty}},
    \label{eq:(p+q).graph-based}
\end{equation}
the $1$-Wasserstein distance between (weak) solutions to~\eqref{eq:Boltz.graph-strong} and~\eqref{eq:Boltz-strong} remains uniformly bounded in time, see Theorem~\ref{theo:exp_relax.graph-based}. Moreover, the graph-based solutions of~\eqref{eq:Boltz.graph-strong} converge exponentially fast in time to the equilibrium distribution arising in the graph-independent setting, see Corollary~\ref{cor:long-time.graph-based.gen}. Consequently, whenever the quantitative condition~\eqref{eq:(p+q).graph-based}, linking the connectivity parameters of the connection network to those of the binary interaction rule, is satisfied, \textit{the influence of the connection network on the aggregate configuration of the system vanishes in the long-time limit}. Since this result applies to a general class of interaction rules, it reveals an intrinsic structural property of the graph-based Boltzmann-type equation~\eqref{eq:Boltz.graph-strong}.

Despite its structural significance, however, condition~\eqref{eq:(p+q).graph-based} may fail to hold for classes of interaction rules commonly employed in the modelling of socio-economic multi-agent systems. This is because, owing to its generality,~\eqref{eq:(p+q).graph-based} does not necessarily capture the specific features of a given interaction rule. A notable example is provided by \textit{compromise-diffusion} rules:
$$ v'=v+\gamma(v_\ast-v)+D(v)\eta, $$
which model exchange dynamics through a deterministic tendency towards consensus, $\gamma(v_\ast-v)$, superimposed with stochastic fluctuations, $D(v)\eta$. Such rules find application in a wide range of contexts, including wealth redistribution and financial transactions~\cite{cordier2005JSP}; knowledge and competence formation~\cite{bisi2026M3AS,pareschi2014PTRSA}; the adoption of new technologies and products~\cite{masali2026AOR,toscani2018PRE}; velocity alignment in flocking, schooling, and swarming of self-propelled particles~\cite{bhattacherjee2015PRE,denk2020PNAS}; vehicular traffic~\cite{tosin2021MCRF}; and consensus-based global particle optimisation~\cite{benfenati2022AMO}.

In Section~\ref{sect:compromise-diffusion} we therefore examine compromise-diffusion rules in detail and establish conditions both on the graphon connectivity and on the interaction parameters -- namely, the coefficient $\gamma$, the Lipschitz constant of $D$, and the variance of the fluctuation $\eta$ -- that ensure convergence of solutions of~\eqref{eq:Boltz.graph-strong} towards the graph-independent equilibrium distributions of~\eqref{eq:Boltz-strong}; see Theorem~\ref{theo:f.to.M}. We also establish that their distance from the transient solutions of~\eqref{eq:Boltz-strong} remains uniformly bounded in time; see Corollary~\ref{cor:exp_relax.graph-based.W2}. To obtain these results, we work with the $2$-Wasserstein distance, which allows us to exploit the specific structure of compromise-diffusion rules more effectively than would be possible within the framework of generic Lipschitz-continuous interaction rules.

Finally, in Section~\ref{sect:compromise-diffusion}, we also prove that, in the \textit{quasi-invariant limit} of weak but frequent compromise-diffusion interactions, solutions to~\eqref{eq:Boltz.graph-strong} converge in the $2$-Wasserstein distance to solutions of the following \textit{graph-based Fokker--Planck-type equation}; see Theorem~\ref{theo:quasi-inv_limit-graph}:
$$ \frac{\partial g}{\partial t}=\gamma\frac{\partial}{\partial v}\left[\left(\cd_W(x)v-\int_0^1\int_\R W(x,x_\ast)v_\ast g(x_\ast,v_\ast,t)\,dv_\ast\,dx_\ast\right)g\right]
        +\frac{\sigma^2\cd_W(x)}{2}\frac{\partial^2}{\partial v^2}\left(D^2(v)g\right). $$
We then use this equation to determine equilibrium distributions explicitly, both in regimes where graph-independent distributions are expected to emerge and in representative cases where the equilibrium distributions retain a dependence on the latent variable $x$; see Proposition~\ref{prop:cont_dep-FP} and Theorem~\ref{theo:FP.graph-based.equil}.

\section{Problem setting}
In this paper, we consider the graph-based homogeneous Boltzmann-type equation~\eqref{eq:Boltz.graph-strong}, where:
\begin{itemize}
\item $v\in\R$ is the agents' microscopic trait, which evolves through binary interactions expressed by a mapping $(\pr{v},\,\pr{v}_\ast)\mapsto (v,\,v_\ast)$. Here, $\pr{v},\,\pr{v}_\ast\in\R$ denote the pre-interaction traits and $v,\,v_\ast\in\R$ the post-interaction traits, while $J>0$ is the Jacobian factor of this transformation;
\item $x\in [0,\,1]$ and $W:[0,\,1]^2\to [0,\,1]$ are a latent variable and a graphon, respectively, describing the structure of the connections between agents. We assume that $W$ is symmetric, i.e. $W(x,x_\ast)=W(x_\ast,x)$ for every $x,\,x_\ast\in [0,\,1]$;
\item $f(x,\cdot,t)\in\cP(\R)$ is the probability distribution of the trait $v$ for agent $x$ at time $t\geq 0$. In particular,
$$ \int_\R f(x,v,t)\,dv=1, \qquad \forall\,x\in [0,\,1],\,t\geq 0. $$
\end{itemize}

Associated with the graphon $W$, we consider the degree function defined in~\eqref{eq:degree}, together with the parameters $\dWinf$, $\norm{\cd_W}{\infty}$ introduced in~\eqref{eq:deg.param}. We will occasionally exploit the symmetry of $W$ to obtain
$$ \int_0^1W(x,x_\ast)\,dx=\cd_W(x_\ast). $$

\begin{remark} \label{rem:lambda_W}
In the particular case when $W$ factors as the product of two functions of a single variable, this implies
$$ W(x,x_\ast)=\frac{\cd_W(x)\cd_W(x_\ast)}{d_W}, \qquad d_W:=\int_0^1\cd_W(x)\,dx, $$
where $d_W\in [0,\,1]$ denotes the average degree of the graphon $W$.

Moreover, when the latent variable $x$ encodes the degree of the vertices, we have $\cd_W(x)\propto x$ and~\eqref{eq:Boltz.graph-strong} reduces to the equation derived in~\cite{nurisso2026EJAM} with $W(x,x_\ast)\propto xx_\ast$. Hence, the theory developed in this paper includes, as a special case, the equations considered in~\cite{nurisso2026EJAM}.
\end{remark}

Equation~\eqref{eq:Boltz.graph-strong} is more conveniently handled in weak form:
\begin{equation}
    \frac{\partial}{\partial t}\int_\R\varphi(v)f(x,v,t)\,dv=\int_0^1\int_\R\int_\R W(x,x_\ast)\ave*{\varphi(v')-\varphi(v)}
        f(x,v,t)f(x_\ast,v_\ast,t)\,dv\,dv_\ast\,dx_\ast,
    \label{eq:Boltz.graph-weak}
\end{equation}
where $\varphi:\R\to\R$ is an arbitrary observable or, in other words, a test function. In this case, $v'$ denotes the post-interaction trait of an agent with pre-interaction trait $v$, the two being linked by a relationship of the form
\begin{equation}
    v'=\Psi(v,v_\ast;\,\omega),
    \label{eq:int_rule}
\end{equation}
where $\Psi:\R\times\R\times\Omega\to\R$ is the interaction function. The latter models how the trait $v$ of an agent changes into $v'$ in consequence of an interaction with another agent with trait $v_\ast$. The relationship~\eqref{eq:int_rule} is called the \textit{interaction rule}. The abstract variable $\omega$ is introduced to account for a possible randomness in the interaction rule. Specifically, it ranges in an abstract sample space $\Omega$ equipped with a $\sigma$-algebra of measurable subsets and a probability measure $P$. The symbol $\ave{\cdot}$ in~\eqref{eq:Boltz.graph-strong},~\eqref{eq:Boltz.graph-weak} denotes expectation with respect to $P$, hence
$$ \ave{\varphi(v')}:=\int_\Omega\varphi(\Psi(v,v_\ast;\,\omega))\,dP(\omega). $$

Even when not explicitly stated, throughout the paper, we consider an interaction function $\Psi$ satisfying the following:
\begin{assumption} \label{ass:Phi.Lipschitz}
For every $\omega\in\Omega$, the function $\Psi$ in~\eqref{eq:int_rule} is Lipschitz continuous with respect to the variables $v,\,v_\ast$. Hence, there exist possibly random coefficients $p,\,q\geq 0$, $p=p(\omega)$ and $q=q(\omega)$, such that
$$ \abs{\Psi(v,v_\ast;\,\omega)-\Psi(w,w_\ast;\,\omega)}\leq p(\omega)\abs{v-w}+q(\omega)\abs{v_\ast-w_\ast} $$
for every $(v,\,v_\ast),\,(w,\,w_\ast)\in\R^2$. In particular, we assume that $p,\,q$ are not simultaneously zero.
\end{assumption}

For future reference, we observe that Assumption~\ref{ass:Phi.Lipschitz} implies that $\Psi$ grows at most linearly with respect to $v,\,v_\ast$:
\begin{equation}
    \abs{\Psi(v,v_\ast;\,\omega)}\leq\abs{\Psi(0,0;\,\omega)}+p(\omega)\abs{v}+q(\omega)\abs{v_\ast},
        \qquad \forall\,v,\,v_\ast\in\R,\ \omega\in\Omega.
    \label{eq:Psi.linear_growth}
\end{equation}

In the sequel, we will omit the variable $\omega$ in $\Psi$, $p$, and $q$ for brevity.

Most of our results will involve a comparison between the solutions of~\eqref{eq:Boltz.graph-strong},~\eqref{eq:Boltz.graph-weak} and those of the classical homogeneous Boltzmann-type equation~\eqref{eq:Boltz-strong} \textit{without graph structure}, which in weak form writes
\begin{equation}
    \frac{d}{dt}\int_\R\varphi(v)F(v,t)\,dv=\int_\R\int_\R\ave*{\varphi(v')-\varphi(v)}F(v,t)F(v_\ast,t)\,dv\,dv_\ast
    \label{eq:Boltz-weak}
\end{equation}
for an arbitrary observable $\varphi$, where $v'$ is again given by~\eqref{eq:int_rule}.

From the modelling perspective, the crucial difference between~\eqref{eq:Boltz.graph-strong}-\eqref{eq:Boltz.graph-weak} and~\eqref{eq:Boltz-strong}-\eqref{eq:Boltz-weak} lies in the interaction rate. In the latter case, the agents in any randomly sampled pair can interact in the unit time. In the former case, instead, they can interact, on average, with probability $W(x,x_\ast)$, namely the probability that they are connected.

\section{Structural asymptotic behaviour of graph-based solutions}
\label{sect:struct_asympt.graph-based}
In this section, we show that, in general, the long-time behaviour of the solutions to~\eqref{eq:Boltz.graph-strong} is structurally analogous to that of the solutions to~\eqref{eq:Boltz-strong}. More precisely, under a basic, albeit possibly restrictive, assumption on the Lipschitz constants $p,\,q$ of $\Psi$, and essentially no assumptions on the graphon $W$, the solutions to~\eqref{eq:Boltz.graph-strong} remain close to those of~\eqref{eq:Boltz-strong} and converge to the same equilibrium distribution as $t\to+\infty$.

A fundamental tool to carry out this analysis is the \textit{Wasserstein distance} between probability measures. Given $s\in\N\setminus\{0\}$ and $\mu,\,\nu\in\cP_s(\R)$, where
$$ \cP_s(\R):=\left\{\mu\in\cP(\R)\,:\,\int_\R\abs{v}^s\,d\mu(v)<+\infty\right\}, $$
the $s$-Wasserstein distance between $\mu$ and $\nu$ is defined as
$$ \cW_s(\mu,\nu):={\left(\min_{\pi\in\Pi(\mu,\,\nu)}\iint_{\R^2}\abs{v-w}^s\,d\pi(v,w)\right)}^{1/s}, $$
where $\pi\in\cP(\R^2)$ is a \textit{coupling} of $\mu,\,\nu$, i.e. a probability measure on the product space $\R^2=\R\times\R$ whose marginals are $\mu,\,\nu$, and $\Pi(\mu,\,\nu)\subseteq\cP(\R^2)$ denotes the set of all possible couplings of $\mu,\,\nu$.

A coupling $\pi^\star\in\Pi(\mu,\,\nu)$ is said to be \textit{optimal} if
$$ \pi^\star\in\operatorname*{arg\,min}\limits_{\pi\in\Pi(\mu,\,\nu)}\iint_{\R^2}\abs{v-w}^s\,d\pi(v,w), $$
in such a way that
$$ \cW_s(\mu,\nu)={\left(\iint_{\R^2}\abs{v-w}^s\,d\pi^\star(v,w)\right)}^{1/s}. $$

Wasserstein distances are a classical tool for optimal transport problems~\cite{ambrosio2008BOOK,villani2009BOOK}. Recently, they have been extended to deal with more general settings in which mass conservation may fail; see~\cite{piccoli2014ARMA,piccoli2016ARMA}. For their application to kinetic equations, the interested reader is referred to~\cite{carrillo2007RMUP,iacobelli2022ARMA} and references therein.

\subsection{Long-time behaviour of graph-independent solutions}
We preliminarily establish some basic facts about the long-time behaviour of the solutions to the classical homogeneous Boltzmann-type equation without graph structure~\eqref{eq:Boltz-strong}.

Let $F^0=F^0(v)$ be the initial condition, i.e. $F(v,0)=F^0(v)$.
\begin{proposition} \label{prop:m_s.bounded-F}
If $F^0\in\cP_s(\R)$ then $F(\cdot,t)\in\cP_s(\R)$ for every $t>0$.
\end{proposition}
\begin{proof}
We take $\varphi(v)=\abs{v}^s$ in~\eqref{eq:Boltz-weak} to get:
\begin{align*}
    \dfrac{d}{dt}\int_\R\abs{v}^sF(v,t)\,dv+\int_\R\abs{v}^sF(v,t)\,dv &=
        \int_\R\int_\R\ave{\abs{\Psi(v,v_\ast)}^s}F(v,t)F(v_\ast,t)\,dv\,dv_\ast, \\
\intertext{whence, invoking~\eqref{eq:Psi.linear_growth} together with Jensen's inequality,}
    &\leq 3^{s-1}\left(\ave{\abs{\Psi(0,0)}^s}+\ave{p^s+q^s}\int_\R\abs{v}^sF(v,t)\,dv\right).
\end{align*}
Multiplying both sides by $e^t$ and integrating on the interval $[0,\,t]$, with $t>0$, yields:
\begin{align*}
    e^t\int_R\abs{v}^sF(v,t)\,dv &\leq \int_\R\abs{v}^sF^0(v)\,dv \\
    &\phantom{\leq} +3^{s-1}\ave{\abs{\Psi(0,0)}^s}\left(e^t-1\right)
        +3^{s-1}\ave{p^s+q^s}\int_0^t e^\tau\int_\R\abs{v}^sF(v,\tau)\,dv\,d\tau.
\end{align*}
Applying now Gr\"{o}nwall's inequality as in~\cite[Lemma~6.2]{loy2026RMUP} to the function $t\mapsto e^t\int_\R\abs{v}^sF(v,t)\,dv$ produces
\begin{align*}
    \int_\R\abs{v}^sF(v,t)\,dv &\leq \left(\int_\R\abs{v}^sF^0(v)\,dv\right)e^{\left(3^{s-1}\ave{p^s+q^s}-1\right)t} \\
    &\phantom{\leq} +3^{s-1}\ave{\abs{\Psi(0,0)}^s}\int_0^te^{\left(3^{s-1}\ave{p^s+q^s}-1\right)(t-\tau)}\,d\tau.
\end{align*}
This shows that $\int_\R\abs{v}^sF(v,t)\,dv$ is finite for every $t>0$, whence the thesis follows.
\end{proof}

Since solutions to~\eqref{eq:Boltz-strong} belong to $\cP_s(\R)$, we can establish a continuous dependence estimate for them in the $s$-Wasserstein metric, which also provides insights into their long-time behaviour. To this end, we employ the \textit{coupling method}, introduced in~\cite{fournier2020EMS-SMS} to study the decay of Wasserstein distances between pairs of solutions to a range of evolution equations.

\begin{proposition} \label{prop:cont_dep.graph-independent}
Let $F(\cdot,t),\,G(\cdot,\,t)\in\cP_s(\R)$ be weak solutions to~\eqref{eq:Boltz-strong} starting from $F^0,\,G^0\in\cP_s(\R)$, respectively. Then
$$ \cW_s(F(t),G(t))\leq\cW_s(F^0,G^0)e^{\frac{1}{s}\left(\ave{(p+q)^s}-1\right)t}, \qquad \forall\,t>0. $$
In particular, if $\ave{(p+q)^s}<1$ then $\cW_s(F(t),G(t))\to 0$ as $t\to +\infty$.
\end{proposition}
\begin{proof}
Let $\pi_t\in\Pi(F(t),\,G(t))$ be a coupling of $F(t),\,G(t)$ starting from an optimal coupling $\pi_0\in\Pi(F^0,\,G^0)$ of $F^0,\,G^0$. We consider the following evolution equation for $\pi_t$:
$$ \frac{d}{dt}\iint_{\R^2}\Phi(v,w)\,d\pi_t(v,w)=\iint_{\R^2}\iint_{\R^2}\ave{\Phi(v',w')-\Phi(v,w)}\,d\pi_t(v,w)\,d\pi_t(v_\ast,w_\ast), $$
where $\Phi:\R^2\to\R$ is an arbitrary observable. Since this is a Boltzmann-type equation on the product space $\R^2$, its solution $\pi_t$ remains a probability measure for all $t>0$. Moreover, both marginals of $\pi_t$ satisfy~\eqref{eq:Boltz-weak}, as can be readily verified by first taking $\Phi$ to depend only on $v$ and then only on $w$.

Let now $\Phi(v,w)=\abs{v-w}^s$. We have:
\begin{align*}
    \resizebox{\textwidth}{!}{$
    \begin{aligned}[b]
    \frac{d}{dt}\iint_{\R^2}\abs{v-w}^s\,d\pi_t(v,w) &= \iint_{\R^2}\iint_{\R^2}
        \ave{\abs{\Psi(v,v_\ast)-\Psi(w,w_\ast)}^s-\abs{v-w}^s}\,d\pi_t(v,w)\,d\pi_t(v_\ast,w_\ast),
\intertext{whence, invoking Assumption~\ref{ass:Phi.Lipschitz},}
    &\leq \iint_{\R^2}\iint_{\R^2}\ave{\left(p\abs{v-w}+q\abs{v_\ast-w_\ast}\right)^s}\,d\pi_t(v,w)\,d\pi_t(v_\ast,w_\ast) \\
    &\phantom{\leq} -\iint_{\R^2}\abs{v-w}^s\,d\pi_t(v,w) \\
    &= \sum_{k=1}^{s}\ave{p^kq^{s-k}}\left(\iint_{\R^2}\abs{v-w}^k\,d\pi_t(v,w)\right)
        \left(\iint_{\R^2}\abs{v_\ast-w_\ast}^{s-k}\,d\pi_t(v_\ast,w_\ast)\right) \\
    &\phantom{=} -\iint_{\R^2}\abs{v-w}^s\,d\pi_t(v,w).
    \end{aligned}
    $}
\end{align*}

H\"{o}lder's inequality with conjugate exponents $\pp=s/k$ and $\qq=s/(s-k)$ yields
$$ \iint_{\R^2}\abs{v-w}^k\,d\pi_t(v,w)\leq\left(\iint_{\R^2}\abs{v-w}^s\,d\pi_t(v,w)\right)^{k/s} $$
and
$$ \iint_{\R^2}\abs{v_\ast-w_\ast}^{s-k}\,d\pi_t(v_\ast,w_\ast)\leq\left(\iint_{\R^2}\abs{v_\ast-w_\ast}^s\,d\pi_t(v_\ast,w_\ast)\right)^{(s-k)/s}. $$
Hence,
$$ \left(\iint_{\R^2}\abs{v-w}^k\,d\pi_t(v,w)\right)
        \left(\iint_{\R^2}\abs{v_\ast-w_\ast}^{s-k}\,d\pi_t(v_\ast,w_\ast)\right)\leq\iint_{\R^2}\abs{v-w}^s\,d\pi_t(v,w) $$
and the previous estimate becomes
$$ \frac{d}{dt}\iint_{\R^2}\abs{v-w}^s\,d\pi_t(v,w)\leq (\ave{(p+q)^s}-1)\iint_{\R^2}\abs{v-w}^s\,d\pi_t(v,w). $$

Consequently,
\begin{align*}
    \iint_{\R^2}\abs{v-w}^s\,d\pi_t(v,w) &\leq \left(\iint_{\R^2}\abs{v-w}^s\,d\pi_0(v,w)\right)e^{(\ave{(p+q)^s}-1)t} \\
    &= \cW_s^s(F^0,G^0)e^{(\ave{(p+q)^s}-1)t},
\end{align*}
where the last equality follows from the fact that $\pi_0$ is optimal. Taking the minimum over all the couplings in $\Pi(F(t),\,G(t))$ on the left-hand side completes the proof.
\end{proof}

In the parameter regime $\ave{(p+q)^s}<1$, Proposition~\ref{prop:cont_dep.graph-independent} implies that there can exist at most one equilibrium distribution $M\in\cP_s(\R)$ for~\eqref{eq:Boltz-strong}. If such an equilibrium distribution exists, it serves as the analogue of the Maxwellian distribution in classical kinetic theory, and every solution to~\eqref{eq:Boltz-strong} converges to $M$ as $t\to +\infty$.

The following result addresses the existence of $M$.

\begin{proposition} \label{prop:M.graph-independent}
If $\ave{(p+q)^s}<1$ there exists a unique equilibrium distribution $M\in\cP_s(\R)$ to~\eqref{eq:Boltz-strong}.
\end{proposition}
\begin{proof}
Since any equilibrium distribution $M$ is constant in time, it results $\frac{\partial M}{\partial t}=0$. Therefore, from~\eqref{eq:Boltz-strong} we deduce that $M$ satisfies
$$ M(v)=\int_\R\ave*{\frac{1}{J}M(\pr{v})M(\pr{v}_\ast)}\,dv_\ast, $$
i.e. it is a fixed point of the gain operator
$$ Q^+(F,F)(v):=\int_\R\ave*{\frac{1}{J}F(\pr{v})F(\pr{v}_\ast)}\,dv_\ast, \qquad F\in\cP_s(\R). $$
In weak form, the above fixed point equation writes
$$ \int_\R\varphi(v)M(v)\,dv=\int_\R\varphi(v)Q^+(M,M)(v)\,dv $$
for every observable $\varphi$, where, owing to~\eqref{eq:Boltz-weak},
\begin{equation}
    \int_\R\varphi(v)Q^+(F,F)(v)\,dv=\int_\R\ave{\varphi(v')}F(v)F(v_\ast)\,dv\,dv_\ast.
    \label{eq:Q+}
\end{equation}
Letting $\varphi\equiv 1$, we see that $Q^+(F,F)$ is a probability distribution for every $F\in\cP_s(\R)$, indeed $\int_\R Q^+(F,F)(v)\,dv=1$; the non-negativity follows from $\int_\R\varphi(v)Q^+(F,F)(v)\,dv\geq 0$ for every $\varphi\geq 0$.

To prove the existence and uniqueness of such a fixed point, we rely on Banach fixed-point theorem in the metric space $(\cP_s(\R),\,\cW_s)$, which is complete.

First, we show that $Q^+$ maps $\cP_s(\R)$ into itself. For this, let $F\in\cP_s(\R)$, then
\begin{align*}
    \int_\R\abs{v}^sQ^+(F,F)(v)\,dv &= \int_\R\int_\R\ave{\abs{\Psi(v,v_\ast)}^s}F(v)F(v_\ast)\,dv\,dv_\ast
\intertext{whence, from~\eqref{eq:Psi.linear_growth} and Jensen's inequality,}
    &\leq 3^{s-1}\left(\ave{\abs{\Psi(0,0)}^s}+\ave{p^s+q^s}\int_\R\abs{v}^sF(v)\,dv\right)<+\infty.
\end{align*}

Now, we show that $Q^+(F,F)$ is a contraction on $\cP_s(\R)$. Let $F,\,G\in\cP_s(\R)$ and let $\pi^\star\in\Pi(F,\,G)$ be an optimal coupling of them. Moreover, let $\pi_{Q^+}\in\Pi(Q^+(F,F),\,Q^+(G,G))$ be the coupling of $Q^+(F,F),\,Q^+(G,G)$ defined by
$$ \iint_{\R^2}\Phi(v,w)\,d\pi_{Q^+}(v,w):=\iint_{\R^2}\iint_{\R^2}\ave{\Phi(v',w')}\,d\pi^\star(v,w)\,d\pi^\star(v_\ast,w_\ast) $$
for every observable $\Phi:\R^2\to\R$. Notice that the marginals of $\pi_{Q^+}$ satisfy~\eqref{eq:Q+}. For $\Phi(v,w)=\abs{v-w}^s$ this gives:
\begin{align*}
    \resizebox{\textwidth}{!}{$
    \begin{aligned}[b]
    \iint_{\R^2}\abs{v-w}^s\,d\pi_{Q^+}(v,w) &= \iint_{\R^2}\iint_{\R^2}\ave{\abs{\Psi(v,v_\ast)-\Psi(w,w_\ast)}^s}\,d\pi^\star(v,w)\,d\pi^\star(v_\ast,w_\ast) \\
    &\leq \iint_{\R^2}\iint_{\R^2}\ave{(p\abs{v-w}+q\abs{v_\ast-w_\ast})^s}\,d\pi^\star(v,w)\,d\pi^\star(v_\ast,w_\ast) \\
    &=\sum_{k=0}^{s}\ave{p^kq^{s-k}}\left(\iint_{\R^2}\abs{v-w}^k\,d\pi^\star(v,w)\right)\left(\iint_{\R^2}\abs{v_\ast-w_\ast}^{s-k}\,d\pi^\star(v_\ast,w_\ast)\right),
\intertext{whence, applying H\"{o}lder's inequality like in the proof of Proposition~\ref{prop:cont_dep.graph-independent} and invoking the optimality of $\pi^\star$,}
    &\leq \ave{(p+q)^s}\cW_s^s(F,G).
    \end{aligned}
    $}
\end{align*}

Minimising over $\pi_{Q^+}\in\Pi(Q^+(F,F),\,Q^+(G,G))$ on the left-hand side shows that $Q^+$ is indeed a contraction in $(\cP_s(\R),\,\cW_s)$ under the assumption $\ave{(p+q)^s}<1$.
\end{proof}

Combining Propositions~\ref{prop:cont_dep.graph-independent},~\ref{prop:M.graph-independent}, we conclude that, whenever $\ave{(p+q)^s}<1$,~\eqref{eq:Boltz-strong} admits a unique equilibrium distribution $M\in\cP_s(\R)$, which is globally asymptotically stable.

\subsection{Exponential relaxation of graph-based solutions to graph-in\-de\-pen\-dent solutions}
We now consider the solutions to~\eqref{eq:Boltz.graph-strong} and show that, if condition~\eqref{eq:(p+q).graph-based} mentioned in Section~\ref{sect:intro} is satisfied, they relax towards the solutions to~\eqref{eq:Boltz-strong}. In particular, they converge to the equilibrium distribution $M$ of Proposition~\ref{prop:M.graph-independent} as $t\to +\infty$.

We regard this as a structural feature of the graph-based Boltzmann-type equation~\eqref{eq:Boltz.graph-strong}, since it holds under the sole assumption~\eqref{eq:(p+q).graph-based}, which, for $s=1$, may naturally be viewed as the graph-based counterpart of the condition in Propositions~\ref{prop:cont_dep.graph-independent},~\ref{prop:M.graph-independent}. In particular, no further assumptions on either the graphon $W$ or the interaction function $\Psi$ are required.

\begin{remark}
Henceforth, we write $f(t)\in L^\infty([0,\,1];\,\cP_s(\R))$ if, for $t\geq 0$, $f(x,\cdot,t)\in\cP_s(\R)$ for almost every $x\in [0,\,1]$ and the map $x\mapsto \int_\R\abs{v}^sf(x,v,t)\,dv$ belongs to $L^\infty(0,\,1)$; that is,
$$ \esup_{x\in [0,\,1]}\int_\R\abs{v}^sf(x,v,t)\,dv<+\infty. $$

For $\mu,\,\nu\in L^\infty([0,\,1];\,\cP_s(\R))$, we measure their distance by the quantity
$$ \esup_{x\in [0,\,1]}\cW_s(\mu(x,\cdot),\nu(x,\cdot)). $$
\end{remark}

Let $f^0=f^0(x,v)$ be the initial condition prescribed to~\eqref{eq:Boltz.graph-strong}, i.e. $f(x,v,0)=f^0(x,v)$.

\begin{proposition} \label{prop:m_s.bounded-f}
If $f^0\in L^\infty([0,\,1];\,\cP_s(\R))$ then $f(t)\in L^\infty([0,\,1];\,\cP_s(\R))$ for every $t>0$.
\end{proposition}
\begin{proof}
Let
$$ \cm_s(x,t):=\int_\R\abs{v}^sf(x,v,t)\,dv. $$
Taking $\varphi(v)=\abs{v}^s$ in~\eqref{eq:Boltz.graph-weak} gives
\begin{align*}
    \frac{\partial\cm_s}{\partial t}(x,t) &= \int_0^1\int_\R\int_\R W(x,x_\ast)\ave{\abs{\Psi(v,v_\ast)}^s-\abs{v}^s}f(x,v,t)f(x_\ast,v_\ast,t)\,dv\,dv_\ast\,dx_\ast,
\intertext{whence, invoking~\eqref{eq:Psi.linear_growth} together with Jensen's inequality,}
    &\leq 3^{s-1}\left[\cd_W(x)\Bigl(\ave{\abs{\Phi(0,0)}^s}+\ave{p^s}\cm_s(x,t)\Bigr)+\ave{q^s}\int_0^1 W(x,x_\ast)\cm_s(x_\ast,t)\,dx_\ast\right] \\
    &\phantom{\leq} -\cd_W(x)\cm_s(x,t) \\
    &\leq 3^{s-1}\cd_W(x)\Bigl[\ave{\abs{\Phi(0,0)}^s}+\ave{p^s+q^s}\norm{\cm_s(t)}{\infty}\Bigr]-\cd_W(x)\cm_s(x,t).
\end{align*}

Moving the term $-\cd_W(x)\cm_s(x,t)$ to the left-hand side, multiplying both sides by $e^{\cd_W(x)t}$, and integrating then over the interval $[0,\,t]$, with $t>0$, produces
$$ e^{\cd_W(x)t}\cm_s(x,t)\leq\cm_s^0(x)+3^{s-1}\norm{\cd_W}{\infty}\int_0^te^{\cd_W(x)\tau}\Bigl[\ave{\abs{\Phi(0,0)}^s}+\ave{p^s+q^s}\norm{\cm_s(\tau)}{\infty}\Bigr]\,d\tau, $$
where $\cm_s^0(x):=\int_\R\abs{v}^sf^0(x,v)\,dv$. Multiplying both sides by $e^{-\cd_W(x)t}$, observing that the right-hand side satisfies $e^{-\cd_W(x)t}\leq e^{-\dWinf t}$ and $e^{\cd_W(x)(\tau-t)}\leq e^{\dWinf(\tau-t)}$, and then multiplying through by $e^{\dWinf t}$, we obtain
$$ e^{\dWinf t}\norm{\cm_s(t)}{\infty}\leq\norm{\cm_s^0}{\infty}
    +3^{s-1}\norm{\cd_W}{\infty}\int_0^te^{\dWinf\tau}\Bigl[\ave{\abs{\Phi(0,0)}^s}+\ave{p^s+q^s}\norm{\cm_s(\tau)}{\infty}\Bigr]\,d\tau. $$
Finally, applying Gr\"{o}nwall's inequality to the function $t\mapsto e^{\dWinf t}\norm{\cm_s(t)}{\infty}$ as in~\cite[Lemma~6.2]{loy2026RMUP}, yields
\begin{align*}
    \norm{\cm_s(t)}{\infty} &\leq \norm{\cm_s^0}{\infty}e^{\left(3^{s-1}\ave{p^s+q^s}\norm{\cd_W}{\infty}-\dWinf\right)t} \\
    &\phantom{\leq} +3^{s-1}\norm{\cd_W}{\infty}\ave{\abs{\Psi(0,0)}^s}\int_0^te^{\left(3^{s-1}\ave{p^s+q^s}\norm{\cd_W}{\infty}-\dWinf\right)(t-\tau)}\,d\tau,
\end{align*}
which shows that $\norm{\cm_s(t)}{\infty}<+\infty$ for every $t>0$, whence the thesis follows.
\end{proof}

Proposition~\ref{prop:m_s.bounded-f} implies, in particular, that, in the stated assumptions, $f(x,\cdot,t)\in\cP_1(\R)$ for almost every $x\in [0,\,1]$ and every $t>0$. Therefore, we can estimate the distance between $f(x,\cdot,t)$ and its graph-independent counterpart $F(\cdot,t)$ by means of the $1$-Wasserstein distance $\cW_1$. This is provided by the following
\begin{theorem} \label{theo:exp_relax.graph-based}
Let $f$ be a weak solution to~\eqref{eq:Boltz.graph-strong} starting from $f^0\in L^\infty([0,\,1];\,\cP_1(\R))$. Likewise, let $F$ be a weak   solution to~\eqref{eq:Boltz-strong} starting from $F^0\in\cP_1(\R)$. Moreover, let $M\in\cP_1(\R)$ be the equilibrium distribution given by Proposition~\ref{prop:M.graph-independent}. Then
\begin{align*}
    \esup_{x\in [0,\,1]}\cW_1(f(x,t),F(t)) &\leq \esup_{x\in [0,\,1]}\cW_1(f^0(x),F^0)e^{\left(\ave{p+q}\norm{\cd_W}{\infty}-\dWinf\right)t} \\
    &\phantom{\leq} +\ave{p+q+1}\norm{1-\cd_W}{\infty}\psi(t)\cW_1(F^0,M)
\end{align*}
for every $t>0$, where the function $\psi$ is defined as
\begin{equation*}
    \psi(t):=
    \begin{cases}
        \dfrac{e^{(\ave{p+q}-1)t}-e^{(\ave{p+q}\norm{\cd_W}{\infty}-\dWinf)t}}
            {\ave{p+q}(1-\norm{\cd_W}{\infty})+\dWinf-1} & \text{if }\ave{p+q}\neq\dfrac{1-\dWinf}{1-\norm{\cd_W}{\infty}}, \\[5mm]
        te^{\frac{\norm{\cd_W}{\infty}-\dWinf}{1-\norm{\cd_W}{\infty}}t} &
            \text{if } \ave{p+q}=\dfrac{1-\dWinf}{1-\norm{\cd_W}{\infty}}.
    \end{cases}
\end{equation*}
\end{theorem}
\begin{remark} \label{rem:W1_vs_Ws}
Here, we restrict our attention to the $1$-Wasserstein distance rather than the more general $s$-Wasserstein distance, since we are comparing two structurally different equations, such as~\eqref{eq:Boltz.graph-strong} and~\eqref{eq:Boltz-strong}, for which it is not straightforward to construct a suitable equation governing a coupling of their respective solutions.

The advantage of working with the $1$-Wasserstein distance is that the Kantorovich--Rubinstein duality allows us to avoid considering such couplings explicitly. More precisely, for any $\mu,\,\nu\in\cP_1(\R)$, this duality yields the representation
$$ \cW_1(\mu,\,\nu)=\sup_{\varphi\in\Lip_1(\R)}\int_{\R}\varphi(v)\,d(\nu-\mu)(v), $$
where $\Lip_1(\R)$ denotes the set of Lipschitz continuous functions on $\R$ with Lipschitz constant at most $1$. See~\cite{ambrosio2008BOOK} for details.

It is this dual representation of $\cW_1$ that we will use in the proof of Theorem~\ref{theo:exp_relax.graph-based}.
\end{remark}
\begin{proof}[Proof of Theorem~\ref{theo:exp_relax.graph-based}]
Let $\varphi\in\Lip_1(\R)$ in both~\eqref{eq:Boltz.graph-weak} and~\eqref{eq:Boltz-weak}. Subtracting the two equations yields
\begin{align*}
    \frac{\partial}{\partial t}\int_{\R}&\varphi(v)(F(v,t)-f(x,v,t))\,dv \\
    &= \int_0^1\int_{\R}\int_{\R}W(x,x_\ast)\ave{\varphi(v')}(F(v,t)F(v_\ast,t)-f(x,v,t)f(x_\ast,v_\ast,t))\,dv\,dv_\ast\,dx_\ast \\
    &\phantom{=} +(1-\cd_W(x))\left(\int_{\R}\int_{\R}\ave{\varphi(v')}F(v,t)F(v_\ast,t)\,dv\,dv_\ast-\int_{\R}\varphi(v)F(v,t)\,dv\right) \\
    &\phantom{=} -\cd_W(x)\int_{\R}\varphi(v)(F(v,t)-f(x,v,t))\,dv,
\end{align*}
where we have repeatedly used~\eqref{eq:degree}.

From the proof of Proposition~\ref{prop:M.graph-independent}, we know that the equilibrium distribution $M$ satisfies
$$ \int_{\R}\int_{\R}\ave{\varphi(v')}M(v)M(v_\ast)\,dv\,dv_\ast-\int_{\R}\varphi(v)M(v)\,dv=0 $$
for every observable $\varphi$. Using this identity in the relation above, we rewrite the latter as
\begin{align*}
    \frac{\partial}{\partial t}\int_{\R}&\varphi(v)(F(v,t)-f(x,v,t))\,dv+\cd_W(x)\int_{\R}\varphi(v)(F(v,t)-f(x,v,t))\,dv \\
    &=\int_0^1W(x,x_\ast)\int_{\R}\left(\int_{\R}\ave{\varphi(v')}(F(v,t)-f(x,v,t))\,dv\right)f(x_\ast,v_\ast,t)\,dv_\ast\,dx_\ast \\
    &\phantom{=}+\int_0^1W(x,x_\ast)\int_{\R}\left(\int_{\R}\ave{\varphi(v')}(F(v_\ast,t)-f(x_\ast,v_\ast,t))\,dv_\ast\right)F(v,t)\,dv\,dx_\ast \\
    &\phantom{=}+(1-\cd_W(x))\left[\int_{\R}\left(\int_{\R}\ave{\varphi(v')}(F(v,t)-M(v))\,dv\right)M(v_\ast)\,dv_\ast\right. \\
    &\phantom{=+(1-\cd_W(x))\left[\right.}+\int_{\R}\left(\int_{\R}\ave{\varphi(v')}(F(v_\ast,t)-M(v_\ast))\,dv_\ast\right)F(v,t)\,dv \\
    &\phantom{=+(1-\cd_W(x))\left[\right.}\left.-\int_{\R}\varphi(v)(F(v,t)-M(v))\,dv\right].
\end{align*}

We now observe that, by Assumption~\ref{ass:Phi.Lipschitz}, the mapping $v\mapsto\ave{\varphi(v')}=\ave{\varphi(\Psi(v,v_\ast))}$ is Lipschitz continuous, uniformly in $v_\ast\in\R$. Indeed, for every $v_1,\,v_2\in\R$ and $\varphi\in\Lip_1(\R)$, we have
$$ \abs{\ave{\varphi(\Psi(v_2,v_\ast)}-\ave{\varphi(\Psi(v_1,v_\ast)}}\leq\ave{\abs{\Psi(v_2,v_\ast)-\Psi(v_1,v_\ast)}}
    \leq\ave{p}\abs{v_2-v_1}. $$
Similarly, the mapping $v_\ast\mapsto\ave{\varphi(v')}$ is Lipschitz continuous, uniformly in $v\in\R$, with Lipschitz constant bounded by $\ave{q}$.

Consequently, we have, for instance,
$$ \int_{\R}\ave{\varphi(v')}(F(v,t)-f(x,v,t))\,dv\leq\ave{p}\cW_1(f(x,t),F(t)), $$
and analogous bounds hold for the other terms in the preceding estimate. We may therefore continue the latter as follows:
\begin{align*}
    \frac{\partial}{\partial t}\int_{\R}&\varphi(v)(F(v,t)-f(x,v,t))\,dv+\cd_W(x)\int_{\R}\varphi(v)(F(v,t)-f(x,v,t))\,dv \\
    &\leq \ave{p}\int_0^1W(x,x_\ast)\cW_1(f(x,t),F(t))\,dx_\ast \\
    &\phantom{\leq} +\ave{q}\int_0^1W(x,x_\ast)\cW_1(f(x_\ast,t),F(t))\,dx_\ast \\
    &\phantom{\leq} +\ave{p+q+1}(1-\cd_W(x))\cW_1(F(t),M) \\
    &\leq \ave{p+q}\norm{\cd_W}{\infty}\esup_{x\in [0,\,1]}\cW_1(f(x,t),F(t))+\ave{p+q+1}\norm{1-\cd_W}{\infty}\cW_1(F(t),M).
\end{align*}

At this point, multiplying both sides by $e^{\cd_W(x)t}$, integrating over $[0,\,t]$ for $t>0$, and then arguing as in the proof of Proposition~\ref{prop:m_s.bounded-f}, we obtain
\begin{align*}
    e^{\dWinf t}\esup_{x\in [0,\,1]}\cW_1(f(x,t),F(t)) &\leq \esup_{x\in [0,\,1]}\cW_1(f^0(x),F^0) \\
    &\phantom{\leq} +\ave{p+q}\norm{\cd_W}{\infty}\int_0^te^{\dWinf\tau}\esup_{x\in [0,\,1]}\cW_1(f(x,\tau),F(\tau))\,d\tau \\
    &\phantom{\leq} +\ave{p+q+1}\norm{1-\cd_W}{\infty}\int_0^te^{\dWinf\tau}\cW_1(F(\tau),M)\,d\tau.
\end{align*}
Applying Gr\"{o}nwall's inequality to the function $t\mapsto e^{\dWinf t}\esup_{x\in [0,\,1]}\cW_1(f(x,t),F(t))$ in the form given in~\cite[Lemma~6.2]{loy2026RMUP}, we conclude that
\begin{align*}
    \esup_{x\in [0,\,1]}\cW_1(f(x,t),F(t)) &\leq \esup_{x\in [0,\,1]}\cW_1(f^0(x),F^0)e^{\left(\ave{p+q}\norm{\cd_W}{\infty}-\dWinf\right)t} \\
    &\phantom{\leq} +\ave{p+q+1}\norm{1-\cd_W}{\infty}\int_0^te^{\left(\ave{p+q}\norm{\cd_W}{\infty}-\dWinf\right)(t-\tau)}\cW_1(F(\tau),M)\,d\tau.
\end{align*}
On the other hand, Proposition~\ref{prop:cont_dep.graph-independent} with $G=M$ gives $\cW_1(F(t),M)\leq\cW_1(F^0,M)e^{(\ave{p+q}-1)t}$. Substituting this estimate into the previous inequality and setting
$$ \psi(t):=\int_0^te^{\left(\ave{p+q}\norm{\cd_W}{\infty}-\dWinf\right)(t-\tau)}e^{(\ave{p+q}-1)\tau}\,d\tau $$
finally yields the desired result.
\end{proof}

\begin{corollary} \label{cor:long-time.graph-based.gen}
Let~\eqref{eq:(p+q).graph-based} hold. Under the same assumptions as in Theorem~\ref{theo:exp_relax.graph-based}, the $\cW_1$-distance between any weak solution of~\eqref{eq:Boltz.graph-strong} and any weak solution of~\eqref{eq:Boltz-strong} remains uniformly bounded in time and tends to $0$ as $t\to +\infty$. Moreover, every weak solution of~\eqref{eq:Boltz.graph-strong} converges, as $t\to +\infty$, to the equilibrium distribution $M\in\cP_1(\R)$ given by Proposition~\ref{prop:M.graph-independent} with $s=1$.
\end{corollary}
\begin{proof}
We observe that, under~\eqref{eq:(p+q).graph-based}, we have $\ave{p+q}\neq\frac{1-\dWinf}{1-\norm{\cd_W}{\infty}}$, since $\frac{1-\dWinf}{1-\norm{\cd_W}{\infty}}\geq 1$ whereas $\ave{p+q}<\frac{\dWinf}{\norm{\cd_W}{\infty}}\leq 1$. Consequently,
$$ \psi(t)=\frac{e^{(\ave{p+q}-1)t}-e^{(\ave{p+q}\norm{\cd_W}{\infty}-\dWinf)t}}{\ave{p+q}(1-\norm{\cd_W}{\infty})+\dWinf-1} $$
and the estimate provided by Theorem~\ref{theo:exp_relax.graph-based} specialises as
\begin{multline*}
    \esup_{x\in [0,\,1]}\cW_1(f(x,t),F(t))\leq
        \esup_{x\in [0,\,1]}\cW_1(f^0(x),F^0)e^{\left(\ave{p+q}\norm{\cd_W}{\infty}-\dWinf\right)t} \\
    +\ave{p+q+1}\norm{1-\cd_W}{\infty}\frac{e^{(\ave{p+q}-1)t}-e^{(\ave{p+q}\norm{\cd_W}{\infty}-\dWinf)t}}
                {\ave{p+q}(1-\norm{\cd_W}{\infty})+\dWinf-1}\cW_1(F^0,M).
\end{multline*}

The first statement is then an immediate consequence of the fact that condition~\eqref{eq:(p+q).graph-based} implies both $\ave{p+q}\norm{\cd_W}{\infty}-\dWinf<0$ and $\ave{p+q}-1<0$.

For the second statement, we apply the same estimate with $F^0=M$. Since $M$ is a stationary solution of~\eqref{eq:Boltz-strong}, it follows that $F(t)=M$ for every $t>0$. Hence,
$$ \esup_{x\in [0,\,1]}\cW_1(f(x,t),M)\leq\esup_{x\in [0,\,1]}\cW_1(f^0(x),M)
    e^{\left(\ave{p+q}\norm{\cd_W}{\infty}-\dWinf\right)t}\to 0 $$
as $t\to +\infty$.
\end{proof}

\section{Compromise-diffusion interactions}
\label{sect:compromise-diffusion}
The conclusions of Section~\ref{sect:struct_asympt.graph-based} are rather general and, broadly speaking, show that, under the sole assumption of the relationship~\eqref{eq:(p+q).graph-based} between the Lipschitz constants of the interaction function and the graphon connectivity parameters, the solutions of~\eqref{eq:Boltz.graph-strong} remain close to, and converge asymptotically in time towards, the solutions of~\eqref{eq:Boltz-strong}. Consequently, in the long-time regime, no significant differences are observed between the graph-based and graph-independent solutions. In particular, this result does not rely on any specific assumptions regarding the fine structure of the graphon.

On the other hand, condition~\eqref{eq:(p+q).graph-based} is more restrictive than it may at first appear and, as a result, may not be satisfied even by commonly used interaction rules.

In this section, we focus on \textit{compromise-diffusion interactions}, in which the function $\Psi$ takes the form
\begin{equation}
    \Psi(v,v_\ast)=v+\gamma(v_\ast-v)+D(v)\eta.
    \label{eq:Psi.compr-diff}
\end{equation}
Here, $\gamma\in (0,\,1)$ is a prescribed parameter, $D:\R\to\R_+$ is a Lipschitz continuous function with Lipschitz constant $L_D>0$, and $\eta\in\R$ is a random variable satisfying
\begin{equation}
    \ave{\eta}=0, \qquad \ave{\eta^2}=\sigma^2>0.
    \label{eq:eta}
\end{equation}
We observe that the parameter $\sigma>0$ is the standard deviation of $\eta$ and that $\sigma^2$ is its variance.

In~\eqref{eq:Psi.compr-diff}, the term $\gamma(v_\ast-v)$ models the agents' tendency to compromise, whereas the term $D(v)\eta$ represents a stochastic fluctuation whose intensity depends on the agents' trait through the function $D$.

Interaction functions of the form~\eqref{eq:Psi.compr-diff} have been widely used, for example, in kinetic models of opinion formation, beginning with~\cite{toscani2006CMS}, and of wealth distribution, beginning with~\cite{cordier2005JSP}, to mention just two representative cases.

It is straightforward to verify that, under the above assumptions, the interaction function~\eqref{eq:Psi.compr-diff} satisfies Assumption~\ref{ass:Phi.Lipschitz}:
$$ \abs{\Psi(v,v_\ast)-\Psi(w,w_\ast)}\leq (1-\gamma+L_D\abs{\eta})\abs{v-w}+\gamma\abs{v_\ast-w_\ast},
    \qquad \forall\,(v,\,v_\ast),\,(w,\,w_\ast)\in\R^2, $$
with $p=1-\gamma+L_D\abs{\eta}$ and $q=\gamma$. However, $\ave{p+q}=1+L_D\ave{\abs{\eta}}\geq 1$. Consequently, the analysis developed in Section~\ref{sect:struct_asympt.graph-based} does not provide a characterisation of the long-time behaviour of the solutions to~\eqref{eq:Boltz.graph-strong} corresponding to the interaction function~\eqref{eq:Psi.compr-diff}.

\begin{remark}
The same issue also arises for the more general class of interaction functions of the form
$$ \Psi(v,v_\ast)=v+I(v,v_\ast)+D(v)\eta, $$
where the deterministic term $I(v,v_\ast)$ does not necessarily model a compromise. If $I:\R\times\R\to\R$ is Lipschitz continuous with Lipschitz constants $L_I',\,L_I''>0$, that is,
$$ \abs{I(v,v_\ast)-I(w,w_\ast)}\leq L_I'\abs{v-w}+L_I''\abs{v_\ast-w_\ast},
    \qquad \forall\,(v,\,v_\ast),\,(w,\,w_\ast)\in\R^2, $$
then
$$ \abs{\Psi(v,v_\ast)-\Psi(w,w_\ast)}\leq (1+L_I'+L_D\abs{\eta})\abs{v-w}+L_I''\abs{v_\ast-w_\ast}
    \qquad \forall\,(v,\,v_\ast),\,(w,\,w_\ast)\in\R^2, $$
with $p=1+L_I'+L_D\abs{\eta}$, $q=L_I''$, and again $\ave{p+q}=1+L_I'+L_I''+L_D\ave{\abs{\eta}}\geq 1$.
\end{remark}

In this section, taking~\eqref{eq:Psi.compr-diff} as a prototypical interaction function, we show that a more refined analysis, exploiting the specific structure of $\Psi$ and some features of the graphon $W$, makes it possible to recover a precise characterisation of the long-time behaviour of the solutions to~\eqref{eq:Boltz.graph-strong}, even when the general condition~\eqref{eq:(p+q).graph-based} is not satisfied.

\subsection{Compromise-diffusion without graph structure}
As before, we begin by considering the graph-independent equation~\eqref{eq:Boltz-strong}, fixing $s=2$. This choice enables us to make explicit use of the properties~\eqref{eq:eta} of the stochastic fluctuation $\eta$, especially its vanishing mean, in the computations that follow. By Proposition~\ref{prop:m_s.bounded-F}, if $F^0\in\cP_2(\R)$, then $F(t)\in\cP_2(\R)$ for every $t>0$.

Substituting $\varphi(v)=v$ in~\eqref{eq:Boltz-weak} and using~\eqref{eq:Psi.compr-diff},~\eqref{eq:eta}, we find that the first moment
$$ U(t):=\int_\R vF(v,t)\,dv $$
is conserved in time. Consequently, an analogue of Proposition~\ref{prop:cont_dep.graph-independent} requires, in the present case, to compare solutions of~\eqref{eq:Boltz-strong} with the same first moment.

For the subsequent analysis, the quantity
\begin{equation}
    \cS:=L_D^2\sigma^2-2\gamma(1-\gamma)
    \label{eq:S}
\end{equation}
will play a key role.

\begin{proposition} \label{prop:cont_dep.graph-independent.compr-diff}
Let $F(t),\,G(t)\in\cP_2(\R)$ be two weak solutions of~\eqref{eq:Boltz-strong} starting from $F^0,\,G^0\in\cP_2(\R)$, respectively. Assume, moreover, that
$$ \int_\R vF^0(v)\,dv=\int_\R vG^0(v)\,dv. $$
Then,
$$ \cW_2(F(t),G(t))\leq\cW_2(F^0,G^0)e^{\frac{1}{2}\cS t}, \qquad \forall\,t>0. $$

In particular, if $\cS<0$ then $\cW_2(F(t),G(t))\to 0$ as $t\to +\infty$.
\end{proposition}
\begin{proof}
Let $\pi_t\in\Pi(F(t),\,G(t))$ be a coupling of $F(t)$ and $G(t)$, with initial datum given by an optimal coupling $\pi_0\in\Pi(F^0,\,G^0)$. Substituting $\Phi(v,w)=\abs{v-w}^2$ in the proof of Proposition~\ref{prop:cont_dep.graph-independent}, and using~\eqref{eq:Psi.compr-diff},~\eqref{eq:eta}, we obtain
\begin{align*}
    \frac{d}{dt}\iint_{\R^2}&\abs{v-w}^2\,d\pi_t(v,w) \\
    &= \iint_{\R^2}\iint_{\R^2}\ave{\abs*{(1-\gamma)(v-w)+\gamma(v_\ast-w_\ast)+(D(v)-D(w))\eta}^2}\,d\pi_t(v,w)\,d\pi_t(v_\ast,w_\ast) \\
    &\phantom{=} -\iint_{\R^2}\abs{v-w}^2\,d\pi_t(v,w) \\
    &= (1-\gamma)^2\iint_{\R^2}\abs{v-w}^2\,d\pi_t(v,w)+\gamma^2\iint_{\R^2}\abs{v_\ast-w_\ast}^2\,d\pi_t(v_\ast,w_\ast) \\
    &\phantom{=} +\sigma^2\iint_{\R^2}\abs{D(v)-D(w)}^2\,d\pi_t(v,w) \\
    &\phantom{=} +2\gamma(1-\gamma)\left(\iint_{\R^2}(v-w)\,d\pi_t(v,w)\right)\left(\iint_{\R^2}(v_\ast-w_\ast)\,d\pi_t(v_\ast,w_\ast)\right) \\
    &\phantom{=} -\iint_{\R^2}\abs{v-w}^2\,d\pi_t(v,w).
\intertext{Using the Lipschitz continuity of $D$ to estimate the third term on the right-hand side, and observing that the fourth term vanishes by conservation of the first moment, we finally get}
    &\leq \cS\iint_{\R^2}\abs{v-w}^2\,d\pi_t(v,w),
\end{align*}
whence the claims follow.
\end{proof}

Under condition $\cS<0$, which in the present setting replaces condition $\ave{(p+q)^s}-1<0$ of Section~\ref{sect:struct_asympt.graph-based}, Proposition~\ref{prop:cont_dep.graph-independent.compr-diff} implies that, for each prescribed first moment $U^0\in\R$, there exists at most one equilibrium distribution $M\in\cP_2(\R)$ of~\eqref{eq:Boltz-strong} such that $\int_\R vM(v)\,dv=U^0$. Moreover, every solution of~\eqref{eq:Boltz-strong} with first moment $U^0$ converges to $M$ as $t\to +\infty$.

To establish the existence of such an $M$, we first observe that the set of probability measures in $\cP_2(\R)$ with prescribed first moment is closed in $\cP_2(\R)$ with respect to the topology induced by the $2$-Wasserstein distance. Indeed, convergence in $\cW_2$ implies convergence of the first moments. Hence, the set
$$ \cP_{2,U^0}(\R):=\left\{\mu\in\cP_2(\R)\,:\,\int_\R v\,d\mu(v)=U^0\right\} $$
endowed with the $2$-Wasserstein distance is a complete metric space.

This observation underpins the following result:
\begin{proposition} \label{prop:M.compr-diff}
If $\cS<0$, then for every $U^0\in\R$ there exists a unique equilibrium distribution $M\in\cP_{2,U^0}(\R)$ of~\eqref{eq:Boltz-strong}.
\end{proposition}
\begin{proof}
Borrowing the setting of Proposition~\ref{prop:M.graph-independent}, we first verify that $Q^+$ maps $\cP_{2,U^0}(\R)$ into itself. To this end, in view of Proposition~\ref{prop:M.graph-independent}, it is sufficient to show that, if $F\in\cP_{2,U^0}(\R)$, then $Q^+(F,F)$ has first moment $U^0$:
\begin{align*}
    \int_\R vQ^+(F,F)(v)\,dv &= \int_\R\int_\R\ave{v+\gamma(v_\ast-v)+D(v)\eta}F(v)F(v_\ast)\,dv\,dv_\ast \\
    &= \int_\R vF(v)\,dv=U^0.
\end{align*}

We next prove that $Q^+$ is a contraction on $\cP_{2,U^0}(\R)$. For this, we use the same coupling $\pi_{Q^+}$ as in the proof of Proposition~\ref{prop:M.graph-independent} and choose $\Phi(v,w)=\abs{v-w}^2$ to find:
\begin{align*}
    \iint_{\R^2}&\abs{v-w}^2\,d\pi_{Q^+}(v,w) \\
    &= \iint_{\R^2}\iint_{\R^2}\ave{\abs{(1-\gamma)(v-w)+\gamma(v_\ast-w_\ast)+(D(v)-D(w))\eta}^2}\,d\pi^\star(v,w)\,d\pi^\star(v_\ast,w_\ast).
\intertext{Proceeding like in the proof of Proposition~\ref{prop:cont_dep.graph-independent.compr-diff}, and invoking the optimality of $\pi^\star$, we obtain}
    &\leq (1+\cS)\cW_2^2(F,G).
\end{align*}
We observe that $1+\cS\geq 1-2\gamma(1-\gamma)\geq\frac{1}{2}$, since $2\gamma(1-\gamma)\leq\frac{1}{2}$ for every $\gamma\in [0,\,1]$. Moreover, for $\cS<0$ it results $1+\cS<1$, whence the claim follows.
\end{proof}

\subsection{Compromise-diffusion with graph structure}
We now consider~\eqref{eq:Boltz.graph-strong} together with~\eqref{eq:Psi.compr-diff},~\eqref{eq:eta}, with the aim of deriving a condition that can replace~\eqref{eq:(p+q).graph-based} and ensure the asymptotic relaxation of the graph-based compromise-diffusion solutions towards the equilibrium distribution of the graph-independent problem characterised above.

Throughout the following analysis, we assume that $f^0\in L^\infty([0,\,1];\,\cP_2(\R))$. Proposition~\ref{prop:m_s.bounded-f} then implies that $f(t)\in L^\infty([0,\,1];\,\cP_2(\R))$ for every $t>0$. Consequently, $f(t)\in L^r([0,\,1];\,\cP_2(\R))$ for every $t>0$ and every $r\geq 1$. We will be particularly concerned with the case $r=2$, equipped with the metric
$$ \left(\int_0^1\cW_2^2(\mu(x,\cdot),\nu(x,\cdot))\,dx\right)^\frac{1}{2}, \qquad \mu,\,\nu\in L^2([0,\,1];\,\cP_2(\R)), $$
or, more often, its square.

\subsubsection{First moment evolution}
\label{sect:first_moment}
We begin by studying the evolution of the first moment of the distribution $f$, namely
$$ u(x,t):=\int_\R vf(x,v,t)\,dv. $$
Substituting $\varphi(v)=v$ into~\eqref{eq:Boltz.graph-weak} and using~\eqref{eq:Psi.compr-diff},~\eqref{eq:eta}, we find that $u$ satisfies
\begin{equation}
    \frac{\partial u}{\partial t}=\gamma\int_0^1W(x,x_\ast)(u(x_\ast,t)-u(x,t))\,dx_\ast,
    \label{eq:u}
\end{equation}
which is a particular case of the equations studied in~\cite{bonnet2022M3AS} in the context of consensus formation in opinion dynamics on graphs; see also~\cite{prisant2026TCNS}.

Let
$$ \bar{u}(t):=\int_0^1u(x,t)\,dx $$
denote the average of $u(t)$. Integrating~\eqref{eq:u} with respect to $x\in [0,\,1]$ and exploiting the symmetry of $W$, we find that
$$ \frac{d\bar{u}}{dt}=\gamma\left(\int_0^1\int_0^1W(x,x_\ast)u(x_\ast,t)\,dx_\ast\,dx-\int_0^1\int_0^1W(x_\ast,x)u(x,t)\,dx\,dx_\ast\right)=0, $$
hence $\bar{u}$ is conserved in time. This observation underpins the characterisation of the long-time behaviour of the solutions to~\eqref{eq:u}.

To proceed, it is convenient to introduce the \textit{graph-Laplacian} $\Lap_W:L^2(0,\,1)\to L^2(0,\,1)$, defined by
$$ (\Lap_Wh)(x):=\int_0^1W(x,x_\ast)(h(x)-h(x_\ast))\,dx_\ast, \qquad h\in L^2(0,\,1), $$
which allows us to rewrite equation~\eqref{eq:u} as
\begin{equation}
    \frac{\partial u}{\partial t}=-\gamma\Lap_Wu.
    \label{eq:du/dt=-Lap(u)}
\end{equation}
Since $f(t)\in L^2([0,\,1];\,\cP_2(\R))$, we claim that $u(t)\in L^2(0,\,1)$. Indeed, if we denote by $E(x,t):=\int_\R v^2f(x,v,t)\,dv$ the second moment of $f$, then $\norm{E(t)}{2}<+\infty$ by assumption. Applying the Cauchy--Schwarz inequality first with respect to $v$ and then with respect to $x$, we obtain
\begin{align*}
    \norm{u(t)}{2}^2=\int_0^1u^2(x,t)\,dx &= \int_0^1\left(\int_\R vf(x,v,t)\,dv\right)^2\,dx\leq\int_0^1E(x,t)\,dx\leq\norm{E(t)}{2}<+\infty.
\end{align*}
Consequently, the term $\Lap_Wu$ in~\eqref{eq:du/dt=-Lap(u)} is well-defined.

Now, let $(\cdot,\cdot):L^2(0,\,1)\times L^2(0,\,1)\to\R$ denote the standard inner product on $L^2(0,\,1)$. Observing that $L^2(0,\,1)$ admits the orthogonal decomposition $L^2(0,\,1)=\cC\oplus\cC^\perp$, where
$$ \cC:=\{h\in L^2(0,\,1)\,:\,h \text{ constant}\}, \qquad \cC^\perp:=\{h\in L^2(0,\,1)\,:\,\bar{h}=0\}, $$
from~\cite[Theorem~4.8]{bonnet2022M3AS} we infer that if
\begin{equation}
    \inf_{h\in\cC^\perp}\frac{(\Lap_Wh,\,h)}{\norm{h}{2}^2}>0,
    \label{eq:inf_LW}
\end{equation}
then there exist constants $\alpha,\,\beta>0$ such that the solution of~\eqref{eq:du/dt=-Lap(u)} with initial datum $u^0\in L^2(0,\,1)$ satisfies
\begin{equation}
    \norm{u(t)-\bar{u}^0}{2}\leq\alpha\norm{u^0-\bar{u}^0}{2}e^{-\beta t}, \qquad \forall\,t>0,
    \label{eq:u(t).conv}
\end{equation}
where $\bar{u}^0:=\int_0^1u^0(x)\,dx$. In other words, under condition~\eqref{eq:inf_LW}, the solutions of~\eqref{eq:du/dt=-Lap(u)}, viz.~\eqref{eq:u}, converge exponentially fast in time to their conserved average.

To exploit the estimate~\eqref{eq:u(t).conv} in the present setting, it is essential to identify conditions on the graphon $W$ that ensure~\eqref{eq:inf_LW} holds. We now discuss two representative examples of such conditions.
\begin{enumerate}[label=(\roman*)]
\item A very simple, albeit rather restrictive, condition is the existence of a constant $\Winf>0$ such that
\begin{equation}
    W(x,x_\ast)\geq\Winf, \qquad \text{for a.e. } (x,\,x_\ast)\in [0,\,1]^2.
    \label{eq:Winf}
\end{equation}
Indeed, by the symmetry of $W$,
$$ (\Lap_Wh,\,h)=\frac{1}{2}\int_0^1\int_0^1W(x,x_\ast)(h(x)-h(x_\ast))^2\,dx\,dx_\ast. $$
Thus, for every $h\in\cC^\perp$,
$$ (\Lap_Wh,\,h)\geq\frac{\Winf}{2}\int_0^1\int_0^1(h(x)-h(x_\ast))^2\,dx\,dx_\ast=\Winf\left(\norm{h}{2}^2-\bar{h}^2\right)=\Winf\norm{h}{2}^2, $$
whence
$$ \inf_{h\in\cC^\perp}\frac{(\Lap_Wh,\,h)}{\norm{h}{2}^2}\geq\Winf. $$

\item A weaker structural assumption on $W$ is the existence of a constant $C>0$ such that
\begin{equation}
    \iint_{A\times A^c}W(x,x_\ast)\,dx\,dx_\ast\geq C\Leb{(A)}\Leb{(A^c)}
    \label{eq:W.cut_condition}
\end{equation}
for every Lebesgue measurable subset $A\subseteq [0,\,1]$, where $A^c$ denotes the complement of $A$ in $[0,\,1]$ and $\Leb$ denotes the Lebesgue measure on $\R$. It is immediate that~\eqref{eq:Winf} implies~\eqref{eq:W.cut_condition}, with $C=\Winf$. The fact that~\eqref{eq:W.cut_condition} implies~\eqref{eq:inf_LW} is instead a non-trivial consequence of more general results; see~\cite{khetan2024AIF,lawler1988TAMS} for proofs and further details.
\begin{remark} \label{rem:disconnected_W}
Condition~\eqref{eq:W.cut_condition} expresses the fact that $W$ does not partition the interval $[0,\,1]$ into disconnected components. A prototypical situation in which~\eqref{eq:W.cut_condition} fails is when there exists a measurable subset $A\subseteq (0,\,1)$, with $0<\Leb{(A)}<1$, such that
$$ \iint_{A\times A^c}W(x,x_\ast)\,dx\,dx_\ast=0. $$
In this case, $W\equiv 0$ a.e. on $A\times A^c$, implying that there are no connections between $A$ and $A^c$. This happens, for instance, with the graphon
$$ W(x,x_\ast)=
    \begin{cases}
        1 & \text{if } (x,\,x_\ast)\in [0,\,a]^2\cup (a,\,1]^2 \\
        0 & \text{otherwise},
    \end{cases} $$
where $0<a<1$, taking $A:=[0,\,a]$ and $A^c=(a,\,1]$.
\end{remark}
\end{enumerate}

In view of the above discussion, we henceforth restrict our attention to graphons $W$ for which~\eqref{eq:inf_LW} holds. This class includes, but is not limited to, graphons satisfying~\eqref{eq:Winf} or~\eqref{eq:W.cut_condition}. Consequently, we can assume that the first moment $u$ of the probability distribution $f$ exhibits the asymptotic behaviour~\eqref{eq:u(t).conv}.

\subsubsection{Exponential relaxation towards graph-independent equilibrium distributions}
We now derive an \textit{a priori} estimate for the $2$-Wasserstein distance between solutions of~\eqref{eq:Boltz.graph-strong} under the compromise-diffusion interaction rules~\eqref{eq:Psi.compr-diff}. In the framework considered in Section~\ref{sect:first_moment}, this estimate will then allow us to establish the long-time relaxation of the graph-based dynamics towards the graph-independent equilibrium distributions $M\in\cP_{2,U^0}(\R)$, thereby extending Proposition~\ref{prop:cont_dep.graph-independent.compr-diff} to the graph-based setting.

For this, a preliminary key observation is that every graph-independent equilibrium distribution $M$ given by Proposition~\ref{prop:M.compr-diff} is also an equilibrium distribution of the graph-based Boltzmann-type equation~\eqref{eq:Boltz.graph-strong}. Indeed,
\begin{align*}
    \int_0^1&\int_{\R}\int_{\R}W(x,x_\ast)\ave{\varphi(v')-\varphi(v)}M(v)M(v_\ast)\,dv\,dv_\ast\,dx_\ast \\
    &= \cd_W(x)\int_{\R}\int_{\R}\ave{\varphi(v')-\varphi(v)}M(v)M(v_\ast)\,dv\,dv_\ast \\
    &= 0
\end{align*}
for every observable $\varphi$.

We begin by proving:
\begin{proposition} \label{prop:cont_dep.graph-based.compr-diff}
Let $f_1(t),\,f_2(t)\in L^2([0,\,1];\,\cP_2(\R))$ be two weak solutions of~\eqref{eq:Boltz.graph-strong} starting from $f_1^0,\,f_2^0\in L^\infty([0,\,1];\,\cP_2(\R))$, respectively. Then,
\begin{align*}
    \int_0^1\cW_2^2(f_1(x,t),f_2(x,t))\,dx &\leq \left(\int_0^1\cW_2^2(f_1^0(x),f_2^0(x))\,dx\right)e^{\hat{\cd}\cS t} \\
    &\phantom{\leq} +2\gamma(1-\gamma)\int_0^t\norm{u_1(\tau)-u_2(\tau)}{2}^2e^{\hat{\cd}\cS(t-\tau)}\,d\tau,
\end{align*}
where
\begin{equation}
    \hat{\cd}:=
    \begin{cases}
        \dWinf & \text{if } \cS<0 \\
        \norm{\cd_W}{\infty} & \text{if } \cS\geq 0.
    \end{cases}
    \label{eq:d_hat}
\end{equation}
\end{proposition}
\begin{proof}
Let $\pi_{x,t}\in\Pi(f_1(x,t),\,f_2(x,t))$ be a coupling of $f_1(x,t)$ and $f_2(x,t)$ evolving from an optimal initial coupling $\pi^0_x\in\Pi(f_1^0(x),\,f_2^0(x))$ according to the Boltzmann-type equation
\begin{align*}
    \frac{\partial}{\partial t}\iint_{\R^2}&\Phi(v,w)\,d\pi_{x,t}(v,w) \\
    &= \int_0^1\iint_{\R^2}\iint_{\R^2}W(x,x_\ast)\ave{\Phi(v',w')-\Phi(v,w)}\,d\pi_{x,t}(v,w)\,d\pi_{x_\ast,t}(v_\ast,w_\ast)\,dx_\ast
\end{align*}
for any observable $\Phi$. Note that both marginals of $\pi_{x,t}$ satisfy~\eqref{eq:Boltz.graph-weak}.

Taking $\Phi(v,w)=\abs{v-w}^2$ and using the interaction function~\eqref{eq:Psi.compr-diff}, we obtain
\begin{align*}
    \frac{\partial\cE}{\partial t} &\leq (L_D^2\sigma^2-\gamma(2-\gamma))\cd_W(x)\cE
        +\gamma^2\int_0^1W(x,x_\ast)\cE(x_\ast,t)\,dx_\ast \\
    &\phantom{\leq} +2\gamma(1-\gamma)(u_1(x,t)-u_2(x,t))\int_0^1W(x,x_\ast)(u_1(x_\ast,t)-u_2(x_\ast,t))\,dx_\ast,
\end{align*}
where, for brevity, we have set
\begin{equation}
    \cE(x,t):=\iint_{\R^2}\abs{v-w}^2\,d\pi_{x,t}(v,w).
    \label{eq:cE}
\end{equation}
Integrating both sides over $x\in [0,\,1]$ then gives
\begin{align*}
    \frac{d}{dt}\int_0^1\cE(x,t)\,dx &\leq \hat{\cd}\cS\int_0^1\cE(x,t)\,dx+2\gamma(1-\gamma){\left(\int_0^1\bigl(u_1(x,t)-u_2(x,t)\bigr)\,dx\right)}^2 \\
    &\leq \hat{\cd}\cS\int_0^1\cE(x,t)\,dx+2\gamma(1-\gamma)\norm{u_1(t)-u_2(t)}{2}^2,
\end{align*}
whence, integrating over $[0,\,t]$, for $t>0$, we obtain
\begin{align*}
    \int_0^1\cE(x,t)\,dx &\leq \int_0^1\cE(x,0)\,dx+\hat{\cd}\cS\int_0^t\int_0^1\cE(x,\tau)\,dx\,d\tau \\
    &\phantom{\leq} +2\gamma(1-\gamma)\int_0^t\norm{u_1(\tau)-u_2(\tau)}{2}^2\,d\tau.
\end{align*}
By the optimality of $\pi_x^0$, we have $\cE(x,0)=\cW_2^2(f_1^0(x),f_2^0(x))$; applying Gr\"{o}nwall's inequality, in the form stated in~\cite[Lemma~6.2]{loy2026RMUP}, to the function $t\mapsto\int_0^1\cE_2(x,t)\,dx$ therefore yields
\begin{align*}
    \int_0^1\cE(x,t)\,dx &\leq \left(\int_0^1\cW_2^2(f_1^0(x),f_2^0(x))\,dx\right)e^{\hat{\cd}\cS t} \\
    &\phantom{\leq} +2\gamma(1-\gamma)\int_0^t\norm{u_1(\tau)-u_2(\tau)}{2}^2e^{\hat{\cd}\cS(t-\tau)}\,d\tau.
\end{align*}
The desired result now follows from the fact that $\cE(x,t)\geq\cW_2^2(f_1(x,t),f_2(x,t))$.
\end{proof}

Proposition~\ref{prop:cont_dep.graph-based.compr-diff} and the fact that the equilibrium distributions $M\in\cP_{2,U^0}(\R)$ given by Proposition~\ref{prop:M.compr-diff} are stationary solutions of~\eqref{eq:Boltz.graph-strong} yield the main result of this section:
\begin{theorem} \label{theo:f.to.M}
Assume $\cS<0$, $\dWinf>0$, and let condition~\eqref{eq:inf_LW} hold. Let $f(t)\in L^2([0,\,1];\,\cP_2(\R))$ be a weak solution of~\eqref{eq:Boltz.graph-strong} evolving from an initial condition $f^0\in L^\infty([0,\,1];\,\cP_2(\R))$, and set  
$$ U^0:=\int_0^1\int_{\R}vf^0(x,v)\,dv\,dx. $$
Furthermore, let $M\in\cP_{2,U^0}(\R)$ be the graph-independent equilibrium distribution with first moment $U^0$ given by Proposition~\ref{prop:M.compr-diff}.

Then,
$$ \lim_{t\to +\infty}\int_0^1\cW_2^2(f(x,t),M)\,dx=0 $$
with exponential convergence.
\end{theorem}
\begin{proof}
Applying Theorem~\ref{prop:cont_dep.graph-based.compr-diff} with $f_1=f$ and $f_2=M$ yields
\begin{align*}
    \int_0^1\cW_2^2(f(x,t),M)\,dx &\leq \left(\int_0^1\cW_2^2(f^0(x),M)\,dx\right)e^{\dWinf\cS t} \\
    &\phantom{\leq} +2\gamma(1-\gamma)\int_0^t\norm{u(\tau)-U^0}{2}^2e^{\dWinf\cS(t-\tau)}\,d\tau.
\end{align*}
Note that the assumption $\cS<0$ implies $\hat{\cd}=\dWinf$ in the estimate of Theorem~\ref{prop:cont_dep.graph-based.compr-diff}.

Under condition~\eqref{eq:inf_LW}, we may invoke~\eqref{eq:u(t).conv} to obtain
$$ \norm{u(t)-U^0}{2}^2\leq\alpha^2\norm{u^0-U^0}{2}^2e^{-2\beta t}. $$
Hence, defining
$$ \psi(t):=\int_0^t e^{-2\beta\tau}e^{\dWinf\cS(t-\tau)}\,d\tau=
\begin{cases}
    te^{\dWinf\cS t} & \text{if } 2\beta+\dWinf\cS=0 \\
    \dfrac{e^{-2\beta t}-e^{\dWinf\cS t}}{2\beta+\dWinf\cS} & \text{if } 2\beta+\dWinf\cS\neq 0,
\end{cases} $$
we arrive at
$$ \int_0^1\cW_2^2(f(x,t),M)\,dx\leq\left(\int_0^1\cW_2^2(f^0(x),M)\,dx\right)e^{\dWinf\cS t}
    +2\alpha^2\gamma(1-\gamma)\norm{u^0-U^0}{2}^2\psi(t). $$

Since, for $\dWinf>0$, the right-hand side decays to zero exponentially as $t\to +\infty$, the claim follows.
\end{proof}

Theorem~\ref{theo:f.to.M} establishes that, under the stated assumptions, the solutions of~\eqref{eq:Boltz.graph-strong} converge in $L^2([0,\,1];\,\cP_2(\R))$ to the graph-independent equilibrium distribution of~\eqref{eq:Boltz-strong}. We can exploit this result to obtain an indirect estimate of the distance between the solutions of~\eqref{eq:Boltz.graph-strong} and~\eqref{eq:Boltz-strong}, in the spirit of Theorem~\ref{theo:exp_relax.graph-based} and Corollary~\ref{cor:long-time.graph-based.gen}:
\begin{corollary} \label{cor:exp_relax.graph-based.W2}
Let $f(t)\in L^2([0,\,1];\,\cP_2(\R))$ and $F(t)\in\cP_2(\R)$ be the weak solutions of~\eqref{eq:Boltz.graph-strong} and~\eqref{eq:Boltz-strong}, respectively, with initial conditions $f^0\in L^\infty([0,\,1];\,\cP_2(\R))$ and $F^0\in\cP_2(\R)$ satisfying
$$ \int_0^1\int_\R vf^0(x,v)\,dv\,dx=\int_\R vF^0(v)\,dv=:U^0. $$
Let also $M\in\cP_{2,U^0}(\R)$ be the graph-independent equilibrium distribution with first moment $U^0$, as given by Proposition~\ref{prop:M.compr-diff}.

Then
\begin{align*}
    \int_0^1\cW_2^2(f(x,t),F(t))\,dx &\leq 2\left[\left(\int_0^1\cW_2^2(f^0(x),M)\,dx\right)e^{\hat{\cd}\cS t}+\cW_2^2(F^0,M)e^{\cS t}\right] \\
    &\phantom{\leq} +4\gamma(1-\gamma)\int_0^t\norm{u(\tau)-U^0}{2}^2e^{\hat{\cd}\cS(t-\tau)}\,d\tau,
\end{align*}
where $\hat{\cd}$ is defined in~\eqref{eq:d_hat} and $u(x,t):=\int_\R vf(x,v,t)\,dv$.

In particular, if $\cS<0$ and condition~\eqref{eq:inf_LW} holds, the distance between $f(t)$ and $F(t)$ remains uniformly bounded in time. Moreover, if $\dWinf>0$, $f(t)$ relaxes towards $F(t)$ as $t\to +\infty$.
\end{corollary}
\begin{proof}
The estimate follows from the triangle inequality, which yields
\begin{align*}
    \cW_2^2(f(x,t),F(t)) &\leq {\bigl[\cW_2(f(x,t),M)+\cW_2(F(t),M)\bigr]}^2 \\
    &\leq 2\left[\cW_2^2(f(x,t),M)+\cW_2^2(F(t),M)\right],
\end{align*}
together with Proposition~\ref{prop:cont_dep.graph-independent.compr-diff}, applied with $G=M$, and Proposition~\ref{prop:cont_dep.graph-based.compr-diff}, applied with $f_1=f$ and $f_2=f_2^0=M$. The final assertions follow directly from Theorem~\ref{theo:f.to.M}.
\end{proof}

We emphasise that, unlike Theorem~\ref{theo:exp_relax.graph-based}, Corollary~\ref{cor:exp_relax.graph-based.W2} does not rely on a direct comparison between the equations governing $f$ and $F$ but instead uses the equilibrium distribution $M$ as a pivot  measure. Indeed, as discussed in Remark~\ref{rem:W1_vs_Ws}, the coupling method does not readily lend itself to estimating the $2$-Wasserstein distance between solutions of structurally different equations, such as~\eqref{eq:Boltz.graph-strong} and~\eqref{eq:Boltz-strong}. The resulting estimate is therefore not sharp: in particular, it does not reflect the case of identical initial conditions for $f$ and $F$. Nevertheless, it establishes a result conceptually analogous to Theorem~\ref{theo:exp_relax.graph-based}, which would otherwise be non-trivial to obtain directly.

\subsection{Graph-based Fokker--Planck asymptotics}
A widely used approach to characterising equilibrium distributions in the kinetic theory of multi-agent systems is the so-called \textit{quasi-invariant limit}, introduced in~\cite{cordier2005JSP,toscani2006CMS} and inspired by the classical grazing collision limit for gas molecules~\cite{degond1992M3AS,landau1936PZSU,villani1998ARMA}.

The idea is to consider a regime of \textit{frequent}, \textit{quasi-invariant} interactions, in which agents undergo numerous small changes in their traits per unit time. This is achieved by scaling the interaction function $\Psi$ by a small parameter $0<\epsilon\ll 1$, so that $v'\approx v$, while simultaneously increasing the interaction rate to $1/\epsilon$. As $\epsilon\to 0^+$, these two effects balance each other, giving rise to a limiting Fokker--Planck-type equation that replaces the Boltzmann-type equation in the kinetic description of the multi-agent system. The resulting Fokker--Planck-type equation often admits explicit steady-state distributions, which describe the asymptotic equilibrium of the system in the quasi-invariant regime.

In this section, we apply this procedure to the graph-based Boltzmann-type equation~\eqref{eq:Boltz.graph-strong} with the compromise-diffusion interaction function~\eqref{eq:Psi.compr-diff} and prove the corresponding quasi-invariant limit in the $2$-Wasserstein distance. For a formal, application-oriented treatment in a similar setting, see~\cite{duering2024JNS}.

We set
\begin{equation}
    v'_\epsilon:=v+\epsilon\gamma(v_\ast-v)+\sqrt{\epsilon}\sigma D(v)Y,
    \label{eq:int_rule.compr-diff.eps}
\end{equation}
where $Y$ is a real-valued centred random variable with $\ave{Y^2}=1$ and $\ave{\abs{Y}^3}<+\infty$. Hence, $\sqrt{\epsilon}\sigma Y$ is a rescaling of $\eta$ in~\eqref{eq:Psi.compr-diff} with variance $\epsilon\sigma^2$. Moreover, we rescale the time variable in~\eqref{eq:Boltz.graph-strong} according to $t\to t/\epsilon$ and introduce the scaled kinetic distribution function
$$ f_\epsilon(x,v,t):=f(x,v,t/\epsilon), $$
which satisfies $\partial_tf_\epsilon=\epsilon^{-1}\partial_t f$. We then rewrite the weak form~\eqref{eq:Boltz.graph-weak} of the Boltzmann-type equation~\eqref{eq:Boltz.graph-strong} in the scaled setting as
\begin{align}
    \begin{aligned}[b]
        \frac{\partial}{\partial t}\int_\R &\varphi(v)f_\epsilon(x,v,t)\,dv \\
        &=\frac{1}{\epsilon}\int_0^1\int_\R\int_\R W(x,x_\ast)\ave*{\varphi(v'_\epsilon)-\varphi(v)}f_\epsilon(x,v,t)f_\epsilon(x_\ast,v_\ast,t)\,dv\,dv_\ast\,dx_\ast.
    \end{aligned}
    \label{eq:Boltz.graph-weak.eps}
\end{align}

Assuming that $\varphi$ is sufficiently smooth, for instance $\varphi\in C^{2,1}(\R)$, and expanding $\varphi(v'_\epsilon)$ in a Taylor series around $v$ to second order, we obtain:
\begin{align}
    \begin{aligned}[b]
        \frac{\partial}{\partial t}\int_\R &\varphi(v)f_\epsilon(x,v,t)\,dv \\
        &= \gamma\int_0^1\int_\R W(x,x_\ast)\varphi'(v)\bigl(u_\epsilon(x_\ast,t)-v\bigr)f_\epsilon(x,v,t)\,dv\,dx_\ast \\
        &\phantom{=} +\frac{\epsilon\gamma^2}{2}\int_0^1\int_\R\int_\R W(x,x_\ast)\varphi''(v)(v_\ast-v)^2f_\epsilon(x,v,t)f_\epsilon(x_\ast,v_\ast,t)\,dv\,dv_\ast\,dx_\ast \\
        &\phantom{=} +\frac{\sigma^2\cd_W(x)}{2}\int_\R\varphi''(v)D^2(v)f_\epsilon(x,v,t)\,dv \\
        &\phantom{=} +\cR_\epsilon(f_\epsilon,f_\epsilon)(x,t),
    \end{aligned}
    \label{eq:Boltz-Taylor}
\end{align}
where
$$ u_\epsilon(x_\ast,t):=\int_\R v_\ast f_\epsilon(x_\ast,v_\ast,t)\,dv_\ast $$
and the remainder $\cR_\epsilon$ is given by
\begin{align*}
    \cR_\epsilon(f_\epsilon,f_\epsilon)(x,t):=\frac{1}{2\epsilon}\int_0^1\int_\R\int_\R W(x,x_\ast)\ave{\bigl(\varphi''(\tilde{v})-\varphi''(v)\bigr)
        &(v'_\epsilon-v)^2} \\
    &\times f_\epsilon(x,v,t)f_\epsilon(x_\ast,v_\ast,t)\,dv\,dv_\ast\,dx_\ast,
\end{align*}
with $\tilde{v}$ lying between $v$ and $v'_\epsilon$. By the Lipschitz continuity of $\varphi''$, we have $\abs{\varphi''(\tilde{v})-\varphi''(v)}\lesssim\abs{v_\epsilon'-v}$, whence
$$ \abs{\cR_\epsilon(f_\epsilon,f_\epsilon)(x,t)}\lesssim\frac{1}{\epsilon}\int_0^1\int_\R\int_\R W(x,x_\ast)\ave{\abs{v'_\epsilon-v}^3}
    f_\epsilon(x,v,t)f_\epsilon(x_\ast,v_\ast,t)\,dv\,dv_\ast\,dx_\ast\sim\pazocal{O}(\epsilon^{1/2}), $$
because $\abs{v'_\epsilon-v}\sim\pazocal{O}(\epsilon^{1/2})$ by~\eqref{eq:int_rule.compr-diff.eps}. Thus, for $\epsilon>0$ small,~\eqref{eq:Boltz-Taylor} is formally close to the following equation:
\begin{align}
    \begin{aligned}[b]
        \frac{\partial}{\partial t}\int_\R\varphi(v)g(x,v,t)\,dv &= \gamma\int_\R\varphi'(v)\bigl(u_W(x,t)-\cd_W(x)v\bigr)g(x,v,t)\,dv \\
        &\phantom{=} +\frac{\sigma^2\cd_W(x)}{2}\int_\R\varphi''(v)D^2(v)g(x,v,t)\,dv,
    \end{aligned}
    \label{eq:FP.graph-weak}
\end{align}
where $g$ has to be understood as a suitable limit of $f_\epsilon$. In~\eqref{eq:FP.graph-weak}, we have denoted
$$ u_W(x,t):=\int_0^1W(x,x_\ast)u(x_\ast,t)\,dx_\ast, $$
where $u$ is the first moment of the distribution function $g$.

In strong form,~\eqref{eq:FP.graph-weak} reads
\begin{equation}
    \frac{\partial g}{\partial t}=\gamma\frac{\partial}{\partial v}\bigl[\bigl(\cd_W(x)v-u_W(x,t)\bigr)g\bigl]
        +\frac{\sigma^2\cd_W(x)}{2}\frac{\partial^2}{\partial v^2}\left(D^2(v)g\right),
    \label{eq:FP.graph-strong}
\end{equation}
which is a Fokker--Planck-type equation with non-constant coefficients that are also parametrised by the latent variable $x$.

The remainder of this section is devoted to proving that the solution $f_\epsilon$ of~\eqref{eq:Boltz.graph-weak.eps} converges to the solution $g$ of~\eqref{eq:FP.graph-weak} as $\epsilon\to 0^+$.

We begin with the following preliminary result.
\begin{lemma} \label{lemma:ueps.Eeps-bound}
For each $\epsilon>0$, let $f_\epsilon(x,t)\in \cP_2(\R)$ be the solution of~\eqref{eq:Boltz.graph-weak.eps} with initial datum $f^0(x)\in\cP_2(\R)$, and let
$$ u_\epsilon(x,t):=\int_\R vf_\epsilon(x,v,t)\,dv, \qquad E_\epsilon(x,t):=\int_\R v^2f_\epsilon(x,v,t)\,dv $$
denote its first and second moments. Then
$$ \int_0^1\abs{u_\epsilon(x,t)}\,dx, \qquad \int_0^1E_\epsilon(x,t)\,dx $$
are bounded uniformly in $\epsilon$ and are integrable on every bounded time interval.
\end{lemma}
\begin{proof}
Taking $\varphi(v)=v$ in~\eqref{eq:Boltz.graph-weak.eps}, we find that $u_\epsilon$ satisfies
$$ \frac{\partial u_\epsilon}{\partial t}+\gamma\cd_W(x)u_\epsilon=\gamma\int_0^1W(x,x_\ast)u_\epsilon(x_\ast,t)\,dx_\ast, $$
which implies
$$ e^{\gamma\cd_W(x)t}\abs{u_\epsilon(x,t)}\leq\abs{u_0(x)}+\gamma\int_0^te^{\gamma\cd_W(x)\tau}\int_0^1W(x,x_\ast)\abs{u_\epsilon(x_\ast,\tau)}\,dx_\ast\,d\tau. $$
Using $\cd_W(x)\geq\dWinf$, as in the proof of Proposition~\ref{prop:m_s.bounded-f}, and integrating over $x\in [0,\,1]$, we obtain
$$ e^{\gamma\dWinf t}\int_0^1\abs{u_\epsilon(x,t)}\,dx\leq\int_0^1\abs{u_0(x)}\,dx
    +\gamma\norm{\cd_W}{\infty}\int_0^te^{\gamma\dWinf\tau}\int_0^1\abs{u_\epsilon(x_\ast,\tau)}\,dx_\ast\,d\tau. $$
Applying Gr\"{o}nwall's inequality to the function $t\mapsto e^{\gamma\dWinf t}\int_0^1\abs{u_\epsilon(x,t)}\,dx$ then yields
$$ \int_0^1\abs{u_\epsilon(x,t)}\,dx\leq\left(\int_0^1\abs{u_0(x)}\,dx\right)e^{\gamma\left(\norm{\cd_W}{\infty}-\dWinf\right)t}. $$
Since the right-hand side is independent of $\epsilon$ and integrable on every bounded time interval, this proves the claim for the first moment.

Similarly, taking $\varphi(v)=v^2$ in~\eqref{eq:Boltz.graph-weak.eps}, we find that $E_\epsilon$ satisfies
\begin{align*}
    \frac{\partial E_\epsilon}{\partial t} &= \gamma(\epsilon\gamma-2)\cd_W(x)E_\epsilon(x,t)+\epsilon\gamma^2\int_0^1W(x,x_\ast)E_\epsilon(x_\ast,t)\,dx_\ast \\
    &\phantom{=} +2\gamma(1-\epsilon)u_\epsilon(x,t)\int_0^1W(x,x_\ast)u_\epsilon(x_\ast,t)\,dx_\ast
        +\sigma^2\cd_W(x)\int_\R D^2(v)f_\epsilon(x,v,t)\,dv.
\end{align*}
For sufficiently small $\epsilon>0$, we have $\epsilon\gamma\leq\gamma$ and $1-\epsilon\leq 1$. Moreover, the non-negativity and Lipschitz continuity of $D$ imply that $D(v)\leq D(0)+L_D\abs{v}$ for all $v\in\R$. Using these bounds and integrating the preceding equation over $x\in [0,\,1]$, we obtain, after rearranging,
\begin{align*}
    \frac{d}{dt}\int_0^1E_\epsilon(x,t)\,dx &\leq\left(L_D^2\sigma^2\norm{\cd_W}{\infty}-2\gamma(1-\gamma)\dWinf\right)\int_0^1E_\epsilon(x,t)\,dx \\ 
    &\phantom{\leq} +2\gamma{\left(\int_0^1\abs{u_\epsilon(x,t)}\,dx\right)}^2 \\
    &\phantom{\leq} +2L_D\sigma^2D(0)\norm{\cd_W}{\infty}\int_0^1\int_\R\abs{v}f_\epsilon(x,v,t)\,dv\,dx \\
    &\phantom{\leq} +\sigma^2D^2(0)\norm{\cd_W}{\infty}.
\end{align*}
Applying the Cauchy--Schwarz inequality first with respect to $v$ and then with respect to $x$, we find that
$$ \int_0^1\int_\R\abs{v}f_\epsilon(x,v,t)\,dv\,dx\leq\int_0^1E_\epsilon^{1/2}(x,t)\,dx\leq{\left(\int_0^1E_\epsilon(x,t)\,dx\right)}^{1/2}. $$
Combining this with the bound on $\int_0^1\abs{u_\epsilon(x,t)}\,dx$ established above gives
\begin{align*}
    \frac{d}{dt}\int_0^1E_\epsilon(x,t)\,dx &\leq\left(L_D^2\sigma^2\norm{\cd_W}{\infty}-2\gamma(1-\gamma)\dWinf\right)\int_0^1E_\epsilon(x,t)\,dx \\ 
    &\phantom{\leq} +2\gamma{\left(\int_0^1\abs{u_0(x)}\,dx\right)}^2e^{2\gamma\left(\norm{\cd_W}{\infty}-\dWinf\right)t} \\
    &\phantom{\leq} +2L_D\sigma^2D(0)\norm{\cd_W}{\infty}{\left(\int_0^1E_\epsilon(x,t)\,dx\right)}^{1/2} \\
    &\phantom{\leq} +\sigma^2D^2(0)\norm{\cd_W}{\infty}.
\end{align*}
Finally, applying $ab\leq\frac{1}{2}(a^2+b^2)$ for $a,\,b\geq 0$ to the third term on the right-hand side, we obtain
\begin{align*}
    \frac{d}{dt}\int_0^1E_\epsilon(x,t)\,dx &\leq\left(L_D^2\sigma^2\norm{\cd_W}{\infty}-2\gamma(1-\gamma)\dWinf+1\right)\int_0^1E_\epsilon(x,t)\,dx \\ 
    &\phantom{\leq} +2\gamma{\left(\int_0^1\abs{u_0(x)}\,dx\right)}^2e^{2\gamma\left(\norm{\cd_W}{\infty}-\dWinf\right)t} \\
    &\phantom{\leq} +\sigma^2D^2(0)\norm{\cd_W}{\infty}\left(2L_D^2\sigma^2\norm{\cd_W}{\infty}+1\right).
\end{align*}
Gr\"{o}nwall's inequality now yields a bound on $\int_0^1E_\epsilon(x,t)\,dx$ that is uniform in $\epsilon$ and integrable on every bounded time interval, whence the claim follows also for the second moment.
\end{proof}

Equipped with Lemma~\ref{lemma:ueps.Eeps-bound}, we are in a position to prove the quasi-invariant limit leading from~\eqref{eq:Boltz.graph-weak.eps} to~\eqref{eq:FP.graph-weak}.
\begin{theorem} \label{theo:quasi-inv_limit-graph}
For each $\epsilon>0$, let $f_\epsilon(t)\in L^2([0,\,1];\,\cP_2(\R))$ be the solution of~\eqref{eq:Boltz.graph-weak.eps} with initial datum $f^0\in L^\infty([0,\,1];\,\cP_2(\R))$. Let $g(t)\in L^2([0,\,1];\,\cP_2(\R))$ be the solution of~\eqref{eq:FP.graph-weak} with the same initial datum $g^0=f^0$. Then, for every $t>0$,
$$ \lim_{\epsilon\to 0^+}\int_0^1\cW_2^2(f_\epsilon(x,t),g(x,t))\,dx=0. $$
\end{theorem}
\begin{proof}
For each $\epsilon>0$, we construct a coupling $\pi_{x,t}^\epsilon\in\Pi(f_\epsilon(x,t),\,g(x,t))$ of the solutions to~\eqref{eq:Boltz.graph-weak.eps} and~\eqref{eq:FP.graph-weak}, starting from an optimal coupling $\pi_x^0\in\Pi(f^0(x),\,f^0(x))$ of their common initial datum. More precisely, we define $\pi_{x,t}^\epsilon$ as the solution of
\begin{align*}
    \frac{\partial}{\partial t}\iint_{\R^2}&\Phi(v,w)\,d\pi_{x,t}^\epsilon(v,w) \\
    &=\frac{1}{\epsilon}\int_0^1\iint_{\R^2}\iint_{\R^2}W(x,x_\ast)\ave{\Phi(v'_\epsilon,w)-\Phi(v,w)}\,d\pi_{x,t}^\epsilon(v,w)\,d\pi_{x_\ast,t}^\epsilon(v_\ast,w_\ast)\,dx_\ast \\
    &\phantom{=} +\gamma\int_0^1\iint_{\R^2}\iint_{\R^2}W(x,x_\ast)\frac{\partial\Phi}{\partial w}(v,w)(w_\ast-w)\,d\pi_{x,t}^\epsilon(v,w)\,d\pi_{x_\ast,t}^\epsilon(v_\ast,w_\ast)\,dx_\ast \\
    &\phantom{=} +\frac{\sigma^2}{2}\int_0^1\iint_{\R^2}W(x,x_\ast)\frac{\partial^2\Phi}{\partial w^2}(v,w)D^2(w)\,d\pi_{x,t}^\epsilon(v,w)\,dx_\ast \\
    &\phantom{=} +\sigma^2\int_0^1\iint_{\R^2}W(x,x_\ast)\frac{\partial^2\Phi}{\partial v\partial w}(v,w)D(v)D(w)\,d\pi_{x,t}^\epsilon(v,w)\,dx_\ast.
\end{align*}
for all observables $\Phi$ on $\R^2$. Since this is a mixed Boltzmann--Fokker--Planck-type equation on the product space $\R^2$, its solution remains a probability measure for all $t>0$, cf.~\cite{fournier2020EMS-SMS}. Moreover, by choosing $\Phi$ to be independent of $w$ or $v$, we see that the first and second marginals of $\pi_{x,t}^\epsilon$ satisfy~\eqref{eq:Boltz.graph-weak.eps} and~\eqref{eq:FP.graph-weak}, respectively, as required.

Choosing $\Phi(v,w)=\abs{v-w}^2$, for which $\partial_w\Phi(v,w)=-2(v-w)$, $\partial^2_{ww}\Phi(v,w)=2$, and $\partial^2_{vw}\Phi(v,w)=-2$, we obtain, after some algebraic rearragement,
\begin{align*}
    \frac{\partial}{\partial t}\iint_{\R^2}&\abs{v-w}^2\,d\pi_{x,t}^\epsilon(v,w) \\
    &= \epsilon\gamma^2\int_0^1\int_\R\int_\R W(x,x_\ast)(v_\ast-v)^2f_\epsilon(x,v,t)f_\epsilon(x_\ast,v_\ast,t)\,dv\,dv_\ast\,dx_\ast \\
    &\phantom{=} +\sigma^2\cd_W(x)\iint_{\R^2}\bigl(D(v)-D(w)\bigr)^2\,d\pi_{x,t}^\epsilon(v,w) \\
    &\phantom{=} -2\gamma\cd_W(x)\iint_{\R^2}\abs{v-w}^2\,d\pi_{x,t}^\epsilon(v,w) \\
    &\phantom{=} +2\gamma\int_0^1\iint_{\R^2}\iint_{\R^2}W(x,x_\ast)(v-w)(v_\ast-w_\ast)\,d\pi_{x,t}^\epsilon(v,w)\,d\pi_{x_\ast,t}^\epsilon(v_\ast,w_\ast)\,dx_\ast.
\intertext{Using the Lipschitz continuity of $D$ and the inequality $ab\leq\frac{1}{2}(a^2+b^2)$ with $a=v-w$ and $b=v_\ast-w_\ast$, we further obtain}
    &\leq \epsilon\gamma^2\int_0^1\int_\R\int_\R W(x,x_\ast)(v_\ast-v)^2f_\epsilon(x,v,t)f_\epsilon(x_\ast,v_\ast,t)\,dv\,dv_\ast\,dx_\ast \\
    &\phantom{=} +(L_D^2\sigma^2-\gamma)\cd_W(x)\iint_{\R^2}\abs{v-w}^2\,d\pi_{x,t}^\epsilon(v,w) \\
    &\phantom{=} +\gamma\int_0^1W(x,x_\ast)\iint_{\R^2}\abs{v_\ast-w_\ast}^2\,d\pi_{x_\ast,t}^\epsilon(v_\ast,w_\ast)\,dx_\ast.
\end{align*}

Integrating over $x\in [0,\,1]$ and recalling the function~\eqref{eq:cE} introduced in the proof of Proposition~\ref{prop:cont_dep.graph-based.compr-diff}, which we now denote by $\cE_\epsilon$ to emphasise its dependence on the scaling parameter $\epsilon$, we arrive at
\begin{align*}
    \frac{d}{dt}\int_0^1\cE_\epsilon(x,t)\,dx &\leq \epsilon\gamma^2\int_0^1\int_0^1W(x,x_\ast)\int_\R\int_\R(v_\ast-v)^2f_\epsilon(x,v,t)f_\epsilon(x_\ast,v_\ast,t)\,dv\,dv_\ast\,dx\,dx_\ast \\
    &\phantom{\leq} +L_D^2\sigma^2\norm{\cd_W}{\infty}\int_0^1\cE_\epsilon(x,t)\,dx,
\end{align*}
where we have repeatedly exploited the symmetry of $W$.

We next observe that
\begin{align*}
    \int_0^1\int_0^1&W(x,x_\ast)\int_\R\int_\R(v_\ast-v)^2f_\epsilon(x,v,t)f_\epsilon(x_\ast,v_\ast,t)\,dv\,dv_\ast\,dx\,dx_\ast \\
    &= 2\left[\int_0^1\cd_W(x)E_\epsilon(x,t)\,dx-\int_0^1\int_0^1W(x,x_\ast)u_\epsilon(x,t)u_\epsilon(x_\ast,t)\,dx\,dx_\ast\right] \\
    &\leq 2\left[\norm{\cd_W}{\infty}\int_0^1E_\epsilon(x,t)\,dx+\left(\int_0^1\abs{u_\epsilon(x,t)}\,dx\right)^2\right],
\end{align*}
where $u_\epsilon$ and $E_\epsilon$ are defined in Lemma~\ref{lemma:ueps.Eeps-bound}. By that lemma, there exists a function $\psi=\psi(t)$, independent of $\epsilon$ and integrable on every bounded time interval, such that
$$ \norm{\cd_W}{\infty}\int_0^1E_\epsilon(x,t)\,dx+\left(\int_0^1\abs{u_\epsilon(x,t)}\,dx\right)^2\leq\frac{\psi(t)}{2}, \qquad \forall\,t>0. $$
Combining the preceding estimates therefore yields
$$ \frac{d}{dt}\int_0^1\cE_\epsilon(x,t)\,dx\leq\epsilon\gamma^2\psi(t)+L_D^2\sigma^2\norm{\cd_W}{\infty}\int_0^1\cE_\epsilon(x,t)\,dx. $$
Since the initial coupling is optimal and both solutions have the same initial datum, we have $\cE_\epsilon(x,0)=0$. Hence, Gr\"{o}nwall's inequality gives
$$ \int_0^1\cE_\epsilon(x,t)\,dx\leq\epsilon\gamma^2\int_0^te^{L_D^2\sigma^2\norm{\cd_W}{\infty}(t-s)}\psi(s)\,ds. $$
But $\cE_\epsilon(x,t)\geq\cW_2^2(f_\epsilon(x,t),g(x,t))$, whence the claim follows.
\end{proof}

\subsubsection{Equilibrium distributions in the quasi-invariant regime}
In this section, we exploit~\eqref{eq:FP.graph-strong} to explicitly characterise equilibrium distributions in the quasi-invariant interaction regime. To this end, we first establish conditions under which solutions to~\eqref{eq:FP.graph-strong} converge to equilibrium as $t\to +\infty$. In doing so, we also identify circumstances in which the equilibrium distributions need not be graph-independent, thus accounting for situations in which the large-time behaviour of the graph-based kinetic equation retains features of the underlying network of connections among agents.

We first observe that the first moment of the distribution function $g$, namely
$$ u(x,t):=\int_\R vg(x,v,t)\,dv, $$
also satisfies~\eqref{eq:u}, as can be readily verified by choosing $\varphi(v)=v$ in~\eqref{eq:FP.graph-weak}. We therefore retain the notation $u$ for this quantity. Consequently, the analysis presented in Section~\ref{sect:first_moment} carries over unchanged to the present setting.

In contrast to the preceding parts of this section, the quantity that will play a crucial role in the analysis that follows is now
$$ \cStilde:=L_D^2\sigma^2-2\gamma. $$
Note that $\cStilde<\cS$, cf.~\eqref{eq:S}.

We begin by establishing a continuous dependence estimate for weak solutions of~\eqref{eq:FP.graph-strong}, in the spirit of Proposition~\ref{prop:cont_dep.graph-based.compr-diff}
\begin{proposition} \label{prop:cont_dep-FP}
Let $g_1(t),\,g_2(t)\in L^2([0,\,1];\,\cP_2(\R))$ be two weak solutions of~\eqref{eq:FP.graph-strong} with respective initial data $g_1^0,\,g_2^0\in L^\infty([0,\,1];\,\cP_2(\R))$. Then, for all $t>0$,
\begin{align*}
    \int_0^1\cW_2^2(g_1(x,t),g_2(x,t))\,dx &\leq \left(\int_0^1\cW_2^2(g_1^0(x),g_2^0(x))\,dx\right)e^{\hat{\cd}\cStilde t} \\
    &\phantom{\leq} +2\gamma\norm{\cd_W}{\infty}^{1/2}\int_0^t\norm{u_1(\tau)-u_2(\tau)}{2}^2e^{\hat{\cd}\cStilde(t-\tau)}\,d\tau,
\end{align*}
where $\hat{\cd}$ is defined in~\eqref{eq:d_hat}.
\end{proposition}
\begin{proof}
Let $\pi_{x,t}\in\Pi(g_1(x,t),\,g_2(x,t))$ be a coupling satisfying
\begin{align*}
    \begin{resizebox}{\textwidth}{!}{$
    \begin{aligned}
    \frac{\partial}{\partial t}&\iint_{\R^2}\Phi(v,w)\,d\pi_{x,t}(v,w) \\
    &= \gamma\iint_{\R^2}\left[\frac{\partial\Phi}{\partial v}(v,w)\bigl(u_{W,1}(x,t)-\cd_W(x)v\bigr)
        +\frac{\partial\Phi}{\partial w}(v,w)\bigl(u_{W,2}(x,t)-\cd_W(x)w\bigr)\right]d\pi_{x,t}(v,w) \\
    &\phantom{=} +\frac{\sigma^2\cd_W(x)}{2}\iint_{\R^2}\left(D^2(v)\frac{\partial^2\Phi}{\partial v^2}(v,w)
        +2D(v)D(w)\frac{\partial^2\Phi}{\partial v\partial w}(v,w)+D^2(w)\frac{\partial^2\Phi}{\partial w^2}(v,w)\right)d\pi_{x,t}(v,w)
    \end{aligned}
    $}
    \end{resizebox}
\end{align*}
for any sufficiently smooth observable $\Phi$, where we have set
$$ u_{W,i}(x,t):=\int_0^1W(x,x_\ast)u_i(x_\ast,t)\,dx_\ast, \qquad i=1,\,2. $$
This is a standard doubling-of-variables argument for Fokker--Planck-type equations, which preserves the non-negativity of the solution $\pi_{x,t}$, cf.~\cite{fournier2020EMS-SMS}. Taking $\Phi$ to be independent of either $v$ or $w$ also shows that both marginals of $\pi_{x,t}$ satisfy~\eqref{eq:FP.graph-weak}, as required. We choose the initial coupling $\pi_x^0\in\Pi(g_1^0(x),\,g_2^0(x))$ to be optimal.

Now, taking $\Phi(v,w)=\abs{v-w}^2$ and noting that $\partial_v\Phi(v,w)=2(v-w)$, $\partial_w\Phi(v,w)=-2(v-w)$, $\partial^2_{vv}\Phi(v,w)=\partial^2_{ww}\Phi(v,w)=2$, $\partial^2_{vw}\Phi(v,w)=-2$, we obtain
\begin{align*}
    \frac{\partial}{\partial t}\iint_{\R^2}\abs{v-w}^2\,d\pi_{x,t}(v,w) &= 2\gamma\bigl(u_{W,1}(x,t)-u_{W,2}(x,t)\bigr)\bigl(u_1(x,t)-u_2(x,t)\bigr) \\
    &\phantom{=} -2\gamma\cd_W(x)\iint_{\R^2}\abs{v-w}^2\,d\pi_{x,t}(v,w) \\
    &\phantom{=} +\sigma^2\cd_W(x)\iint_{\R^2}\left(D(v)-D(w)\right)^2\,d\pi_{x,t}(v,w) \\
    &\leq 2\gamma\bigl(u_{W,1}(x,t)-u_{W,2}(x,t)\bigr)\bigl(u_1(x,t)-u_2(x,t)\bigr) \\
    &\phantom{\leq} +\cStilde\cd_W(x)\iint_{\R^2}\abs{v-w}^2\,d\pi_{x,t}(v,w),
\end{align*}
owing to the Lipschitz continuity of $D$. Recalling the function $\cE$ introduced in the proof of Proposition~\ref{prop:cont_dep.graph-based.compr-diff}, cf.~\eqref{eq:cE}, and integrating the above estimate over $x\in [0,\,1]$, we obtain
$$ \frac{d}{dt}\int_0^1\cE(x,t)\,dx\leq 2\gamma\norm{\cd_W}{\infty}^{1/2}\norm{u_1(t)-u_2(t)}{2}^2+\cStilde\hat{\cd}\int_0^1\cE(x,t)\,dx. $$
Here, we have used the estimate
\begin{align*}
    \int_0^1&\bigl(u_{W,1}(x,t)-u_{W,2}(x,t)\bigr)\bigl(u_1(x,t)-u_2(x,t)\bigr)\,dx \\
    &\leq \norm{u_{W,1}(t)-u_{W,2}(t)}{2}\cdot\norm{u_1(t)-u_2(t)}{2} \\
    &= \left[\int_0^1\left(\int_0^1W(x,x_\ast)\bigl(u_1(x_\ast,t)-u_2(x_\ast,t)\bigr)\,dx_\ast\right)^2\,dx\right]^{1/2}\norm{u_1(t)-u_2(t)}{2} \\
    &\leq \left[\int_0^1\left(\int_0^1W^2(x,x_\ast)\,dx_\ast\right)dx\right]^{1/2}\norm{u_1(t)-u_2(t)}{2}^2 \\
    &\leq \left[\int_0^1\cd_W(x)\,dx\right]^{1/2}\norm{u_1(t)-u_2(t)}{2}^2 \\
    &\leq \norm{\cd_W}{\infty}^{1/2}\norm{u_1(t)-u_2(t)}{2}^2
\end{align*}
which follows from successive applications of the Cauchy--Schwarz inequality and the bound $W^2\leq W$, valid since $0\leq W\leq 1$.

Finally, applying Gr\"{o}nwall's inequality and using $\cE(x,t)\geq\cW_2^2(g_1(x,t),g_2(x,t))$, together with the optimality of $\pi_x^0$, gives the claimed estimate.
\end{proof}

Building on Proposition~\ref{prop:cont_dep-FP}, we can now identify two settings, depending on whether or not the graphon $W$ satisfies condition~\eqref{eq:inf_LW}, in which solutions to~\eqref{eq:FP.graph-strong} may converge to an equilibrium.

\begin{theorem} \label{theo:FP.graph-based.equil}
Under the assumptions of Proposition~\ref{prop:cont_dep-FP}, suppose that $\cStilde<0$, $\dWinf>0$, and that either of the following conditions holds:
\begin{enumerate}[label=(\roman*)]
\item The graphon $W$ satisfies~\eqref{eq:inf_LW} and
$$ \int_0^1\int_\R vg_1^0(x,v)\,dv\,dx=\int_0^1\int_\R vg_2^0(x,v)\,dv\,dx=: \bar{u}^0. $$
\item The initial data $g_1^0$ and $g_2^0$ have the same first moment, i.e.
$$ \int_\R vg_1^0(x,v)\,dv=\int_\R vg_2^0(x,v)\,dv, \qquad \text{for a.e. } x\in [0,\,1] $$
with no further assumptions on the graphon $W$.
\end{enumerate}
Then, in either case,
$$ \lim_{t\to +\infty}\int_0^1\cW_2^2(g_1(x,t),g_2(x,t))\,dx=0. $$
\end{theorem}
\begin{proof}
Since $\cStilde<0$, it follows from~\eqref{eq:d_hat} that $\hat{\cd}=\dWinf$. Hence, the estimate of Proposition~\ref{prop:cont_dep-FP} becomes
\begin{align*}
    \int_0^1\cW_2^2(g_1(x,t),g_2(x,t))\,dx &\leq \left(\int_0^1\cW_2^2(g_1^0(x),g_2^0(x))\,dx\right)e^{\dWinf\cStilde t} \\
    &\phantom{\leq} +2\gamma\norm{\cd_W}{\infty}^{1/2}\int_0^t\norm{u_1(\tau)-u_2(\tau)}{2}^2e^{\dWinf\cStilde(t-\tau)}\,d\tau.
\end{align*}

We consider cases (i) and (ii) separately.
\begin{enumerate}[label=(\roman*)]
\item If $W$ satisfies condition~\eqref{eq:inf_LW} then~\eqref{eq:u(t).conv} applies to both $u_1$ and $u_2$, which are such that
$$ \int_0^1u_1(x,t)\,dx=\int_0^1u_2(x,t)\,dx=\bar{u}^0. $$
Consequently, there exist $\alpha,\,\beta>0$ such that
$$ \max\{\norm{u_1(t)-\bar{u}^0}{2},\,\norm{u_2(t)-\bar{u}^0}{2}\}\leq\alpha C_0e^{-\beta t} $$
for $C_0:=\max\{\norm{u_1^0-\bar{u}^0}{2},\,\norm{u_2^0-\bar{u}^0}{2}\}$ and all $t>0$. It follows that
$$ \norm{u_1(t)-u_2(t)}{2}^2\leq 2(\norm{u_1(t)-\bar{u}^0}{2}^2+\norm{u_2(t)-\bar{u}^0}{2}^2)
        \leq 4\alpha^2C_0^2e^{-2\beta t} $$
for all $t>0$. Substituting this bound into the preceding estimate yields
\begin{align*}
    \int_0^1\cW_2^2(g_1(x,t),g_2(x,t))\,dx &\leq \left(\int_0^1\cW_2^2(g_1^0(x),g_2^0(x))\,dx\right)e^{\dWinf\cStilde t} \\
    &\phantom{\leq} +8\alpha^2C_0^2\gamma\norm{\cd_W}{\infty}^{1/2}\int_0^te^{\dWinf\cStilde(t-\tau)-2\beta\tau}\,d\tau \\
    &= \left(\int_0^1\cW_2^2(g_1^0(x),g_2^0(x))\,dx\right)e^{\dWinf\cStilde t}+8\alpha^2C_0^2\gamma\norm{\cd_W}{\infty}^{1/2}\psi(t),
\end{align*}
where
$$ \psi(t):=
    \begin{cases}
        \dfrac{e^{\dWinf\cStilde t}-e^{-2\beta t}}{\dWinf\cStilde+2\beta} & \text{if } \dWinf\cStilde+2\beta\neq 0 \\[5mm]
        te^{\dWinf\cStilde t} & \text{if } \dWinf\cStilde+2\beta=0.
    \end{cases} $$
Since $\dWinf\cStilde<0$ and $\beta>0$, we have $\psi(t)\to 0$ as $t\to +\infty$ in either case. Therefore, on the whole, the right-hand side of the above estimate is infinitesimal as $t\to +\infty$, whence the claim follows.
\item If $u_1^0(x)=u_2^0(x)$ then, by uniqueness of solutions to~\eqref{eq:u}, cf.~\cite{taricco2026PREPRINT}, we have $u_1(t)=u_2(t)$ for all $t>0$. Hence,
$$ \int_0^1\cW_2^2(g_1(x,t),g_2(x,t))\,dx\leq\left(\int_0^1\cW_2^2(g_1^0(x),g_2^0(x))\,dx\right)e^{\dWinf\cStilde t}\longrightarrow 0 $$
as $t\to +\infty$, and the claim follows also in this case. \qedhere
\end{enumerate}
\end{proof}

Theorem~\ref{theo:FP.graph-based.equil} establishes that, in the stated assumptions, there exists at most one equilibrium distribution of~\eqref{eq:FP.graph-strong}, which is asymptotically stable. We now use~\eqref{eq:FP.graph-strong} to determine the equilibrium distribution explicitly.

Rewriting~\eqref{eq:FP.graph-strong} in divergence form,
$$ \frac{\partial g}{\partial t}=\frac{\partial}{\partial v}\left\{\gamma\bigl[\bigl(\cd_W(x)v-u_W(x,t)\bigr)g\bigl]
        +\frac{\sigma^2\cd_W(x)}{2}\frac{\partial}{\partial v}\left(D^2(v)g\right)\right\}, $$
we obtain the equilibrium distribution $M$ by imposing the zero-flux condition.

In case (i) of Theorem~\ref{theo:FP.graph-based.equil}, we observe that the first moment of $M$ is equal to $\bar{u}^0$, so that $u_W(x)=\cd_W(x)\bar{u}^0$. Consequently, $M$ satisfies
$$ \gamma[(v-\bar{u}^0)M]+\frac{\sigma^2}{2}\frac{\partial}{\partial v}\left(D^2(v)M\right)=0 $$
for all $x\in [0,\,1]$. Here, we have cancelled the factor $\cd_W(x)$, which is non-zero for every $x\in [0,\,1]$ by the assumption $\dWinf>0$ in Theorem~\ref{theo:FP.graph-based.equil}. As expected when the graphon $W$ satisfies~\eqref{eq:inf_LW}, the equilibrium distribution $M$ is independent of $x$ and is given by
\begin{equation}
    M(v)=\frac{1}{D^2(v)}\exp{\left(\frac{2\gamma}{\sigma^2}\int\frac{\bar{u}^0-v}{D^2(v)}\,dv\right)}.
    \label{eq:M.x-homog}
\end{equation}
In this expression, the indefinite integral denotes a primitive of $(\bar{u}^0-v)/D^2(v)$, with the constant of integration chosen so as to satisfy the normalisation condition $\int_\R M(v)\,dv=1$. We note that~\eqref{eq:M.x-homog} coincides with the classical equilibrium distribution of the Fokker--Planck equation arising in graph-independent models, such as those describing wealth redistribution~\cite{cordier2005JSP} and opinion formation~\cite{toscani2006CMS}.

By contrast, in case (ii) of Theorem~\ref{theo:FP.graph-based.equil}, the graphon $W$ need not satisfy~\eqref{eq:inf_LW}, and hence the first moment of $M$ need not be independent of $x$. More precisely, we seek this first moment, denoted by $u^0=u^0(x)$, among the time-independent solutions of~\eqref{eq:u}, so that, taking $g_2=M$ in Theorem~\ref{theo:FP.graph-based.equil}(ii), we can conclude that every weak solution of~\eqref{eq:FP.graph-strong} whose initial condition has first moment $u^0(x)$ converges to $M$ as $t\to +\infty$.

To allow for $x$-dependent time-independent solutions of~\eqref{eq:u}, we consider a setting in which condition~\eqref{eq:inf_LW} is violated as illustrated in Remark~\ref{rem:disconnected_W}. Specifically, we assume that there exists a measurable set $A\subseteq (0,\,1)$, with $0<\Leb{(A)}<1$, such that $W\equiv 0$ on $A\times A^c$, and, by symmetry, also on $A^c\times A$. We then fix two constants $u^0_1,\,u^0_2\in\R$ and show that
\begin{equation}
    u^0(x):=u^0_1\chi_A(x)+u^0_2\chi_{A^c}(x),
    \label{eq:u0.equilibrium}
\end{equation}
where $\chi$ denotes the characteristic function, is a time-independent solution of~\eqref{eq:u}. Indeed, substituting $u^0$ into~\eqref{eq:u} yields
\begin{align*}
    0 &= \int_0^1W(x,x_\ast)(u^0(x_\ast)-u^0(x))\,dx_\ast \\
    &= \int_AW(x,x_\ast)(u^0(x_\ast)-u^0(x))\,dx_\ast+\int_{A^c}W(x,x_\ast)(u^0(x_\ast)-u^0(x))\,dx_\ast \\
    &= \int_AW(x,x_\ast)(u^0_1-u^0(x))\,dx_\ast+\int_{A^c}W(x,x_\ast)(u^0_2-u^0(x))\,dx_\ast.
\end{align*}
For $x\in A$, the first integral on the right-hand side vanishes because $u^0(x)=u_1^0$, while the second vanishes because $W(x,x_\ast)=0$ for $x\in A$ and $x_\ast\in A^c$. Similarly, for $x\in A^c$, the second integral vanishes because $u^0(x)=u^0_2$, while the first vanishes because $W(x,x_\ast)=0$ for $x\in A^c$ and $x_\ast\in A$. Thus, the above identity holds for all $x\in [0,\,1]$.

For $u^0$ given by~\eqref{eq:u0.equilibrium}, we have
$$ u_W(x)=\int_0^1W(x,x_\ast)u^0(x_\ast)\,dx_\ast=u^0_1\int_AW(x,x_\ast)\,dx_\ast+u^0_2\int_{A^c}W(x,x_\ast)\,dx_\ast. $$
Moreover, if $x\in A$, then $\int_AW(x,x_\ast)\,dx_\ast=\cd_W(x)$, since $\int_{A^c}W(x,x_\ast)\,dx_\ast=0$. Likewise, if $x\in A^c$, then $\int_{A^c}W(x,x_\ast)\,dx_\ast=\cd_W(x)$, since $\int_AW(x,x_\ast)\,dx_\ast=0$. Hence,
\begin{align*}
    u_W(x) &=
    \begin{cases}
        \cd_W(x)u^0_1 & \text{if } x\in A \\
        \cd_W(x)u^0_2 & \text{if } x\in A^c
    \end{cases} \\
    &= \cd_W(x)u^0(x).
\end{align*}
Consequently, for $\dWinf>0$, the equilibrium distribution $M$ satisfies
$$ \gamma[(v-u^0(x))M]+\frac{\sigma^2}{2}\frac{\partial}{\partial v}\left(D^2(v)M\right)=0, $$
which gives
$$ M(x,v)=\frac{1}{D^2(v)}\exp{\left(\frac{2\gamma}{\sigma^2}\int\frac{u^0(x)-v}{D^2(v)}\,dv\right)}. $$
In this case, the dependence of $M$ on the latent variable $x$ is explicitly retained through the mean $u^0(x)$ defined in~\eqref{eq:u0.equilibrium}.

\section*{Acknowledgements}
The author is member of GNFM (Gruppo Nazionale per la Fisica Matematica) of INdAM (Istituto Nazionale di Alta Matematica), Italy.

\section*{Data Availability Statement}
No datasets were generated or analysed during the present study.
	
\bibliographystyle{plain}
\bibliography{biblio}
\end{document}